\documentclass[%
 reprint,
 amsmath,amssymb,
 aps,
]{revtex4-1}

\usepackage{graphicx}
\usepackage{epsfig}
\usepackage{epstopdf}
\usepackage{subfigure,color}
\usepackage{dcolumn}
\usepackage{bm}

\begin{document}
	
\preprint{APS/123-QED}

\title{Complementary metasurfaces for guiding electromagnetic waves}

\author{X.~Ma$^{1}$, M.~S.~Mirmoosa$^{2,3}$, and S.~A.~Tretyakov$^{3}$}

\affiliation{$^1$School of Electronics and Information, Northwestern Polytechnical University, 710129 Xi'an, China\\$^2$Laboratory of Wave Engineering, {\'E}cole Polytechnique F{\'e}d{\'e}rale de Lausanne (EPFL), CH-1015 Lausanne, Switzerland\\$^3$Department of Electronics and Nanoengineering, Aalto University, P.O.~Box 15500, FI-00076 Aalto, Finland}

\date{\today}

\begin{abstract}  
Waveguides are critically important components in microwave, THz, and optical technologies. Due to recent progress in two-dimensional materials, metasurfaces can be efficiently used to design novel waveguide structures which confine the electromagnetic energy while the structure is open. Here, we introduce a special type of such structures formed by two penetrable metasurfaces which have {\it complementary} isotropic surface impedances. We theoretically study guided modes supported by the proposed structure and discuss the corresponding dispersion properties. Furthermore, we show the results for different scenarios in which the surface impedances possess non-resonant or resonant characteristics, and the distance between the metasurfaces changes from large values to the extreme limit of zero. As an implication of this work, we demonstrate that there is a possibility to excite two modes with orthogonal polarizations having the same phase velocity within a broad frequency range. This property is promising for applications in leaky-wave antennas and field focusing.
\end{abstract}
\maketitle


\section{Introduction}
Planar structures with periodically arranged subwavelength elements, coined as metasurfaces~\cite{Tretyakov,Holloway,Glybovski,bian_metasurfaces,chenreview}, have attracted considerable attention in the last  decade (see e.g.~\cite{Estakhri,FranciscoCuesta,Eleftheriades101,XWangAalto,mirmoosaaalto,FuAalto,kivshar2015,asadchy2016,yang2014,capasso2011}). One important application of such artificial surfaces is in controlling surface waves~\cite{Tcvetkova,sun,Smith,engheta}, including a single metasurface~\cite{Tretyakov} or a metasurface over a grounded slab~\cite{grbic2,Elek}. This guiding property occurs since we can engineer the surface impedance of such structures~\cite{bilow,hessel,sievenpiper1,sievenpiper2}. Analogous to these planar waveguides, new designs for open waveguide structures can be introduced, which are formed by two parallel and penetrable metasurfaces separated by a finite distance~\cite{xin}. Such open waveguides confine the electromagnetic energy while the corresponding fields are attenuated away from the structure (note that these guiding structures can be intriguingly invisible under plane-wave illumination~\cite{FranciscoCuesta}). An interesting special case of those waveguides is the set of two penetrable metasurfaces which are Babinet-complementary. This structure is intriguing since the product of the surface impedances of two complementary sheets  must be nondispersive and equal to $\eta_0^2/4$, where  $\eta_0$ is the impedance of the background isotropic medium (free space, in this work). Moreover, in the limit of zero distance between the sheets, they combine into a continuous PEC sheet. In this paper, we study this special case of two complementary sheets and investigate the corresponding guiding waves. 

For microwave waveguides, complementary inclusions are realized by interchanging  metal and vacuum regions for a given planar structure, assuming the perfect electric conductor model for metals. Incorporation of such inclusions have been studied and applied also in planar waveguide structures~\cite{Zentgraf,Cui,Kumar,Bitzer}. However, earlier works studied  planar waveguide structures which contain complementary inclusions only in one single layer~\cite{Dong,Landy,Pulido-Mancera,Gonzalez-Ovejero,Sievenpiper3}. Here, we instead impose the condition that the two metasurfaces are complementary with respect to each other. It is worth noting that some attention has been paid to the scattering characteristics of two parallel complementary sheets placed in close proximity of each other~\cite{Lockyer,Bukhari,Bukhari2,our_comp}. However, to the best of our knowledge, there is not enough knowledge on eigenmodes of coupled complementary metasurfaces operating as a single waveguiding structure.

The proposed guiding structure is shown in Fig.~\ref{fig:cs}. Two complementary metasurfaces are separated by the distance $d$, and the space between them is filled by air (vacuum). The surface impedances of the two metasurfaces are denoted as $Z_{\rm{s}1}$ and $Z_{\rm{s}2}$, respectively, and we assume that these values do not depend on the spatial coordinates in the sheet planes. Based on the Babinet principle, we have 
\begin{equation}
Z_{\rm{s}1} \cdot Z_{\rm{s}2}=\frac{\eta_0^2}{4}.
\label{eq:babprin}
\end{equation}
We orient the $z$-axis along the direction of the wave propagation. Here, we classify the proposed structure into two different categories: non-resonant and resonant structures. For each category, we analyze the corresponding guided modes and study the extreme case when the distance between the two metasurfaces tends to zero. Furthermore, we also reveal a possibility of exciting two modes of different polarizations propagating with the same phase velocity (degeneracy state). 

The paper is organized as follows: Sections~\ref{sec:nresis} and \ref{sec:rdis} study two non-resonant and resonant structures. Section~\ref{sec:PI} illustrates the polarization insensitivity, and finally Section~\ref{sec:conclusion} concludes the paper.


\section{Non-resonant dispersive impedance sheets}
\label{sec:nresis} 

\begin{figure*} 
\begin{minipage}[b]{\columnwidth}
\subfigure[]{\includegraphics[width=1.2\columnwidth,height=0.9\columnwidth]{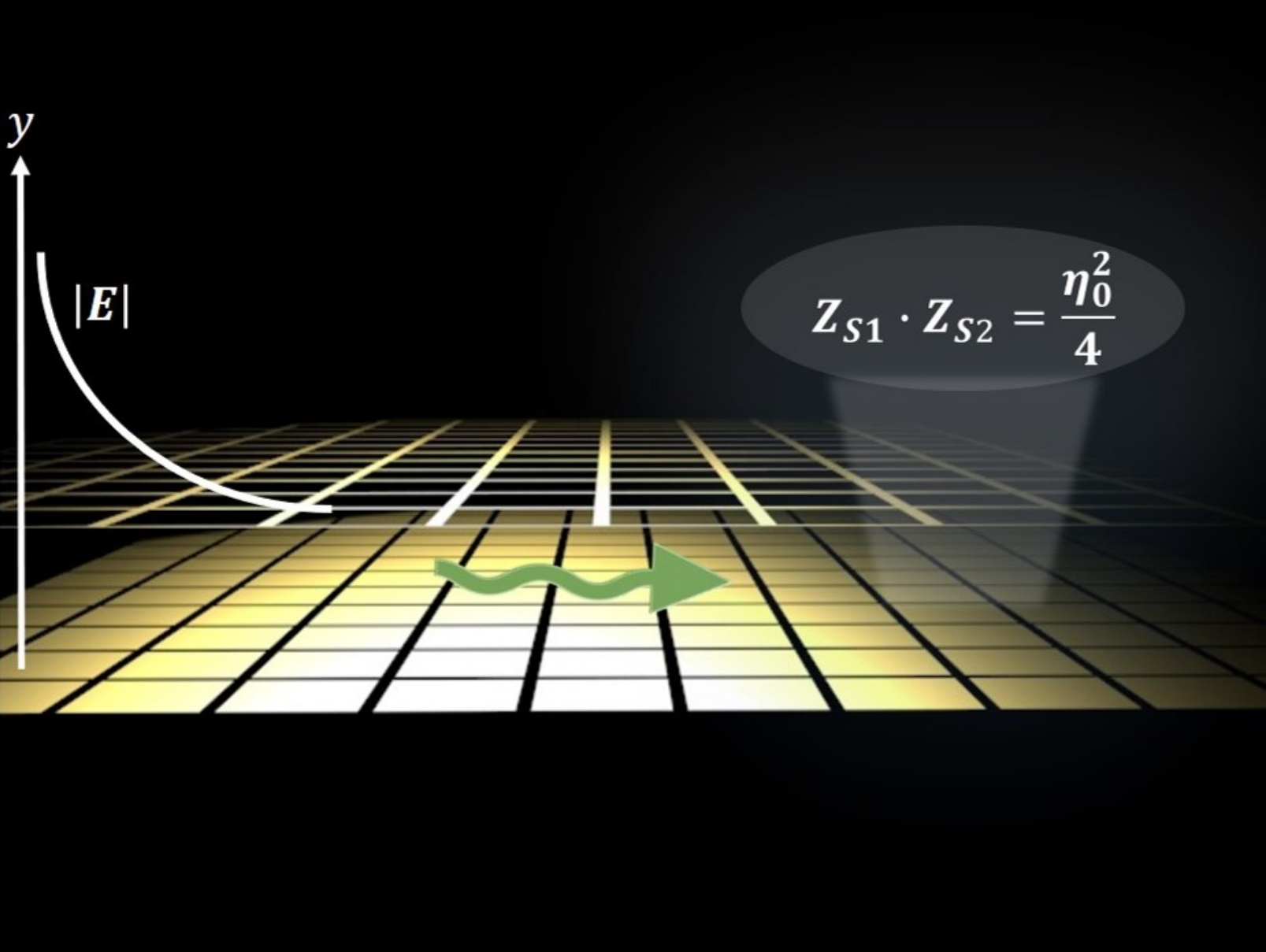}\label{fig:cs}}
\end{minipage}
\hfill
\begin{minipage}[b]{\columnwidth}
\subfigure[]{\includegraphics[width=0.5\columnwidth,height=0.425\columnwidth]{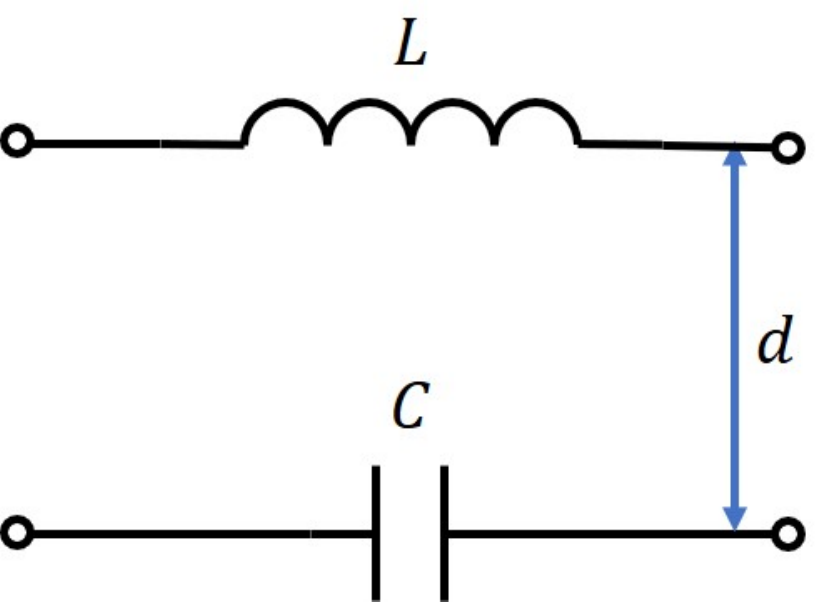}\label{fig:ESnonresonant}}
\subfigure[]{\includegraphics[width=0.5\columnwidth,height=0.425\columnwidth]{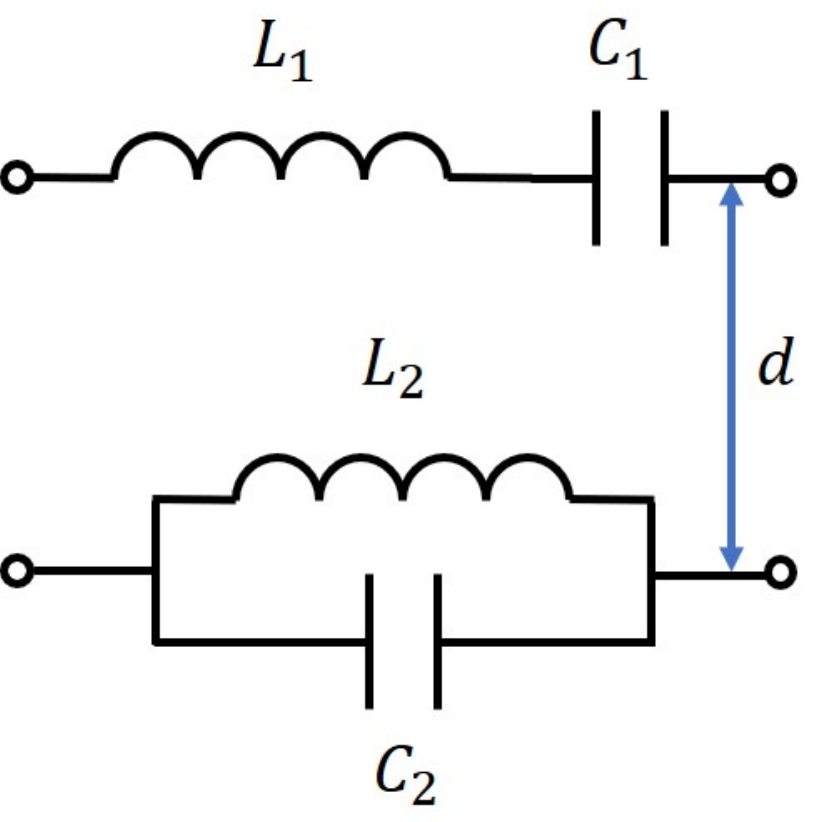}\label{fig:ESresonant}}
\end{minipage}
\caption{(a)--The structure under study: Two complementary impedance sheets. (b)--Equivalent circuit model for two non-resonant dispersive impedance sheets. (c)--Equivalent circuit model for two resonant dispersive impedance sheets.}
\end{figure*}

In this category, the surface reactances are far from resonances (we assume a lossless structure). In other words, they are corresponding to the reactance of a single capacitance or inductance. If the two metasurfaces, shown in Fig.~\ref{fig:cs}, are complementary to each other, one of the metasurfaces should be inductive and the other metasurface should be capacitive. Under this condition, we keep the product of the two surface impedances constant (canceling out the frequency $\omega$) and satisfy the Babinet principle. As a consequence, there are two modes propagating along the waveguide and having transverse magnetic (TM) and transverse electric (TE) polarizations. Note that one single metasurface can support guided waves (surface waves) that possess only transverse electric or transverse magnetic  polarization~\cite{Collin,Tretyakov}. However, in waveguides which consist of two parallel metasurfaces~\cite{xin}, there are always two simultaneous modes whose polarizations depend on the surface impedances. Two inductive (capacitive) sheets correspond to guiding of two TM modes (TE modes), and one inductive and one capacitive sheet correspond to guiding of co-existing TM and TE modes. The equivalent circuit model of two non-resonant dispersive impedance sheets is shown in Fig.~\ref{fig:ESnonresonant}. Practical realization of such structure can be in form of dense meshes of metallic strips for the inductive sheet and arrays of small metallic patches for the capacitive sheet. Let us denote the effective inductance of the inductive sheet by $L$, and the effective capacitance of the other sheet as $C$. From Eq.~\ref{eq:babprin}, we can immediately see that those effective parameters are related to each other as (see also Eq.~(\ref{eq:LC}) of Appendix): 
\begin{equation}
C=\frac{4L}{\eta_0^2}.
\label{eq:cccc}
\end{equation}
Based on the above expression, the dispersion relation of the guided modes can be written in terms of $L$ only: 
\begin{equation}
\frac{\alpha^2}{\epsilon_0}(1-e^{-2\alpha d})+\frac{\alpha}{2L}\eta_0^2=\omega^2(\mu_0+2\alpha L),
\label{eq:TMNonResonantCom}
\end{equation}
for the TM-polarized wave and
\begin{equation}
\frac{\alpha^2}{\epsilon_0}+\frac{\alpha}{2L}\eta_0^2=\omega^2\Big[\mu_0(1-e^{-2\alpha d})+2\alpha L\Big],
\label{eq:TENonResonantCom}
\end{equation}
for the TE-polarized wave. Here $\alpha$ denotes the field attenuation constant for fields outside the waveguide. Contemplating Eq.~\ref{eq:TENonResonantCom}, we see that there is a cut-off frequency for the TE wave, which is associated with the distance between the two metasurfaces. Studying the limit of  $\alpha$ approaching zero, by using l'H{\^{o}}pital's rule we find the cut-off frequency:
\begin{equation}
f_{\rm{cut-off}}={{\frac{\eta_0}{4\pi}}\frac{1}{\sqrt{L(L+\mu_0 d)}}}.
\label{eq:NonResonantComCutOff}
\end{equation}

\begin{figure}[t!]	
	\centerline{\includegraphics[width=0.8\columnwidth]{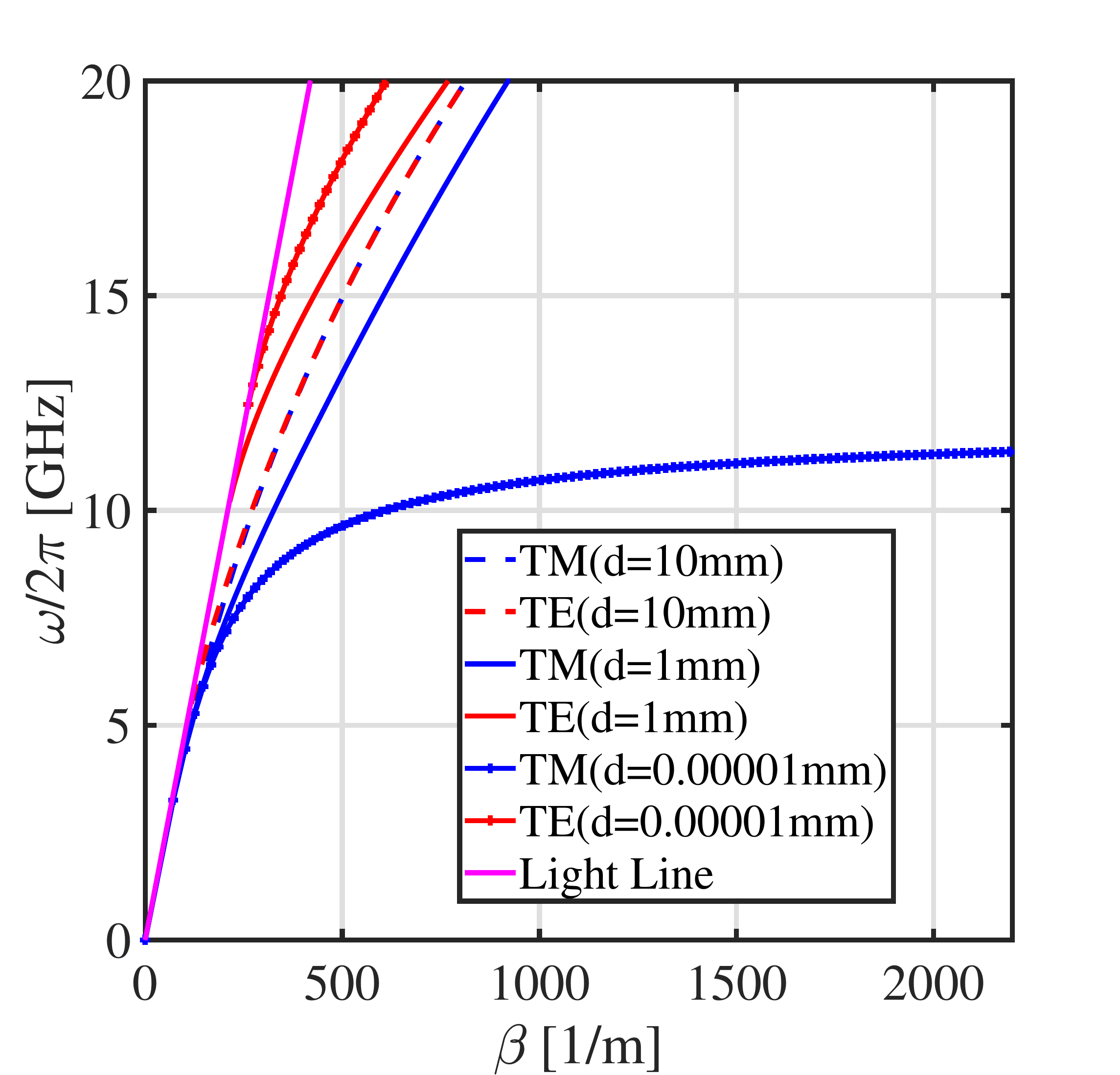}}
	\caption{Dispersion curves of two non-resonant sheets, when the distance between the sheets is $d=10\, {\rm{mm}}$,  $d=1\, {\rm{mm}}$, and  $d=0.00001\, {\rm{mm}}$.}
	\label{fig:NonResonantDispersion}
\end{figure}

\begin{figure}[t!]	
\centerline{\includegraphics[width=0.8\columnwidth]{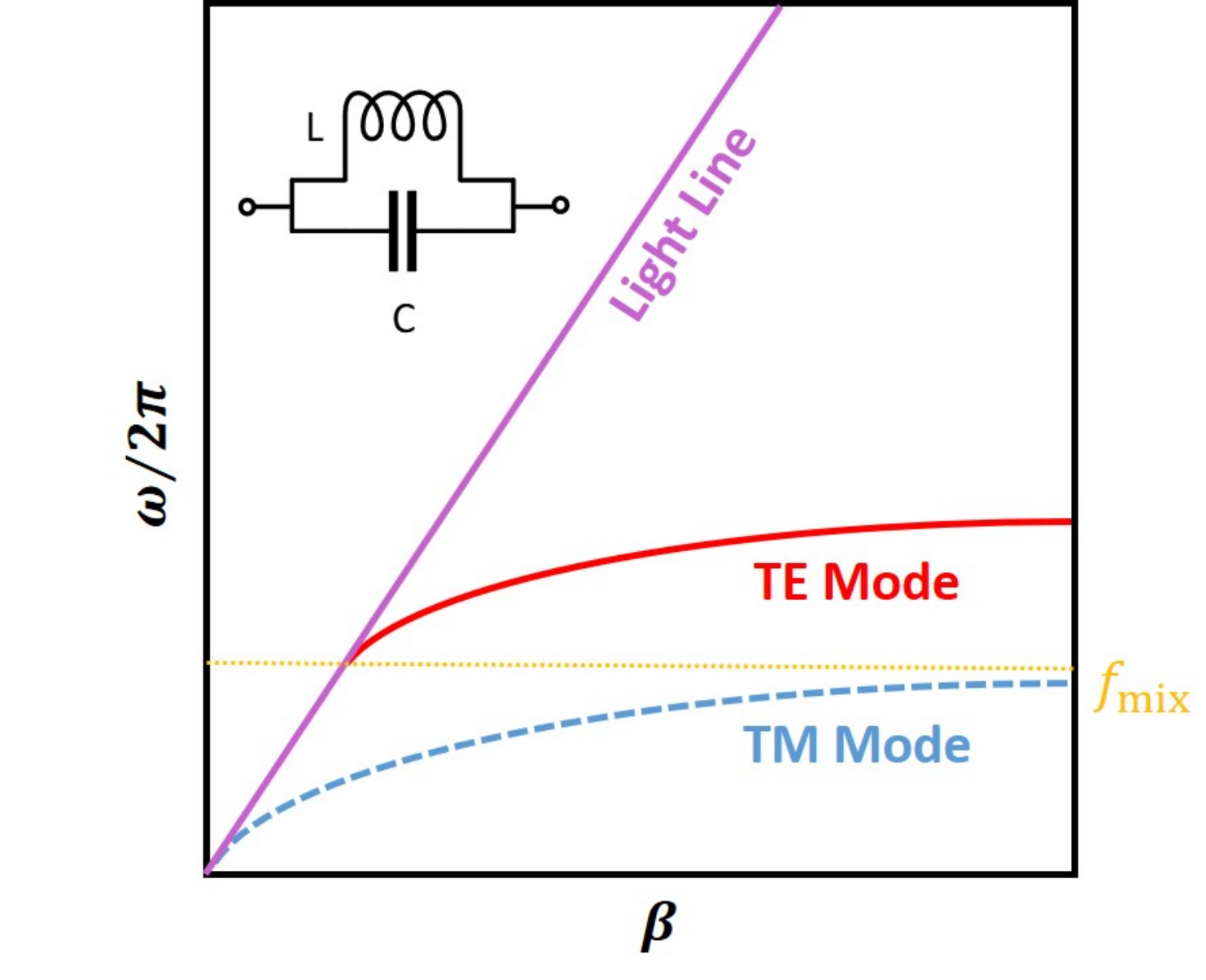}}
\caption{Dispersion curves for two non-resonant sheets in which the distance between two sheets tends to zero.}
\label{fig:NonResonantExtreme}
\end{figure}

As an example, taking $L=2.5$~nH, we can readily find the corresponding value of $C$ in according to Eq.~(\ref{eq:cccc}). Suppose that $\beta$ is the phase constant along the wave propagation direction. Figure~\ref{fig:NonResonantDispersion} shows the analytically calculated dispersion curves (frequency versus $\beta$) for different distances between the two non-resonant sheets. Concurring with expectations, both TM and TE modes can simultaneously propagate along such waveguide. When $d=10$~mm, the TE and TM modes have approximately the same phase velocity within a certain frequency range, which we will explain later in Section~\ref{sec:PI}. Decreasing the distance causes that the dispersion curve corresponding to the TE mode (red line) moves counterclockwise towards the light line at high frequencies and the curve corresponding to the TM mode (blue line) moves clockwise becoming far away from the light line. Interestingly, the limit of $d\rightarrow 0$ leads to the appearance of a new resonance frequency for TM mode, which can be analytically found as
\begin{equation}
f_{\rm{mix}}\approx{{\frac{1}{2\pi}}\frac{1}{\sqrt{LC}}}.
\label{eq:NonResonantfmix}
\end{equation}
This new resonance is exactly the same as the cut-off frequency of the TE-polarized wave (Eq.~(\ref{eq:NonResonantComCutOff})) when $d=0$. The dispersion curve for this extreme case is illustrated in Fig.~\ref{fig:NonResonantExtreme}. 

In order to get better physical insight into the electromagnetic behavior of the waveguide structure, Fig.~\ref{fig:NonResonantfield} plots the electric field distribution, and the following Figs.~\ref{fig:NonResMagCur}~and~\ref{fig:NonResPhaCur} show the surface current density. For the TM mode, the operating frequency is assumed to be $2.7$~GHz and two different distances are considered. For the TE modes, we assume that the operating frequency is $5$~GHz. In Fig.~\ref{fig:NonResonantfield}, note that the top metasurface is inductive and the bottom one is  capacitive. Also, the two black solid lines show the positions of the two metasurfaces. 

One can see that for the TM polarization the field is strongly tied to the inductive sheet and attenuates as the vertical distance from that sheet increases, while the field is strongly confined to the capacitive sheet for the TE polarization. When the two sheets get close to each other, strong coupling can be observed from the phenomenon that the magnitude of electric field does not change too much between the two sheets because of the small distance. The surface current corresponding to the TE-polarized mode flows along the $x$-axis, while it flows along the $z$-axis for the TM polarization. It is perceived that the unequal magnitude of surface current density on the two sheets can be attained due to the asymmetric impedances. Also, as Fig.~\ref{fig:NonResPhaCur} shows, the TM-polarized mode is characterized by out-of-phase surface currents. In contrast, for the  TE-polarized mode, the surface currents are in phase.

\begin{figure}[t!]\centering
	\subfigure[]{\includegraphics[width=0.11\textwidth]{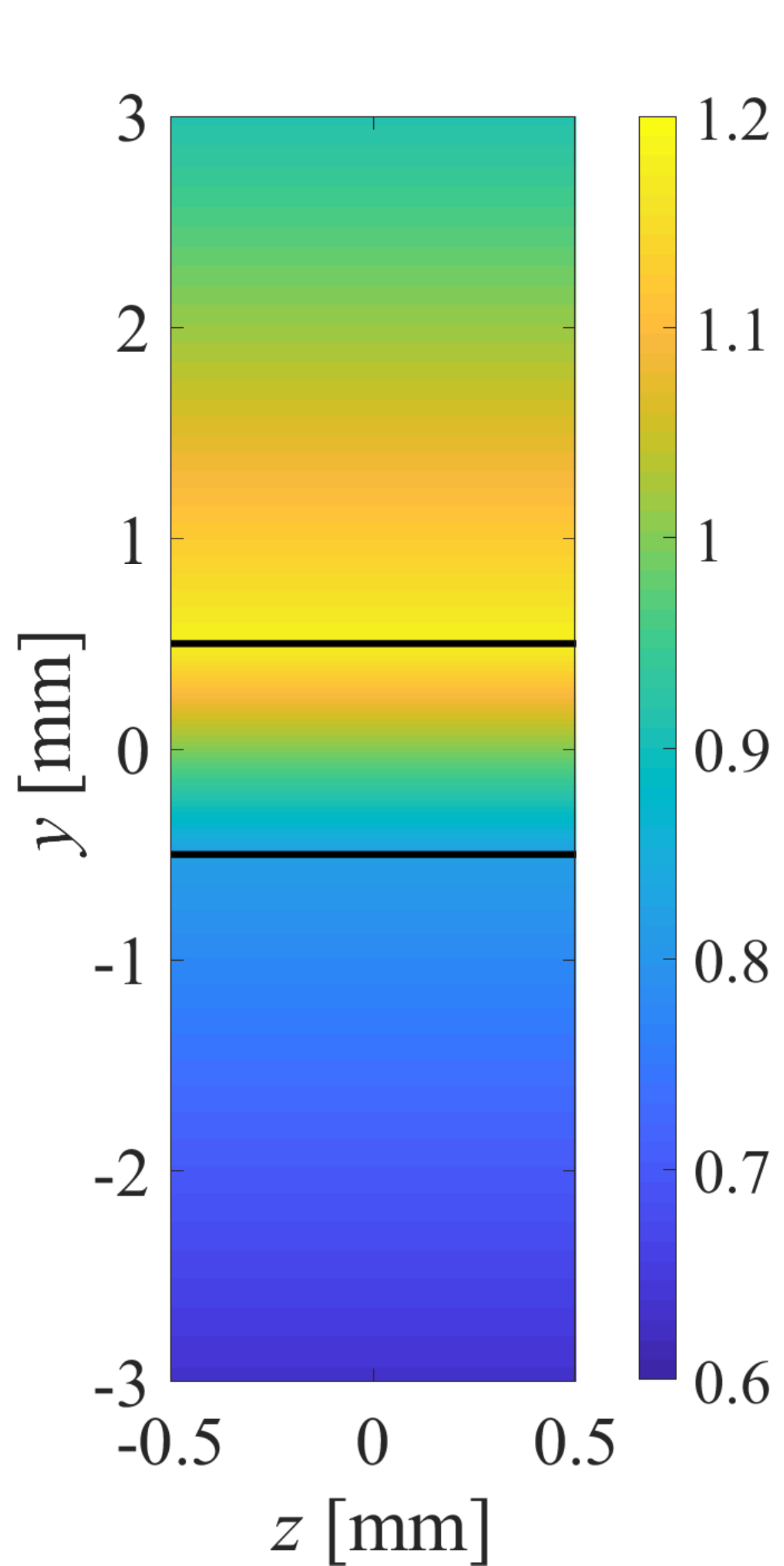}}
	\subfigure[]{\includegraphics[width=0.11\textwidth]{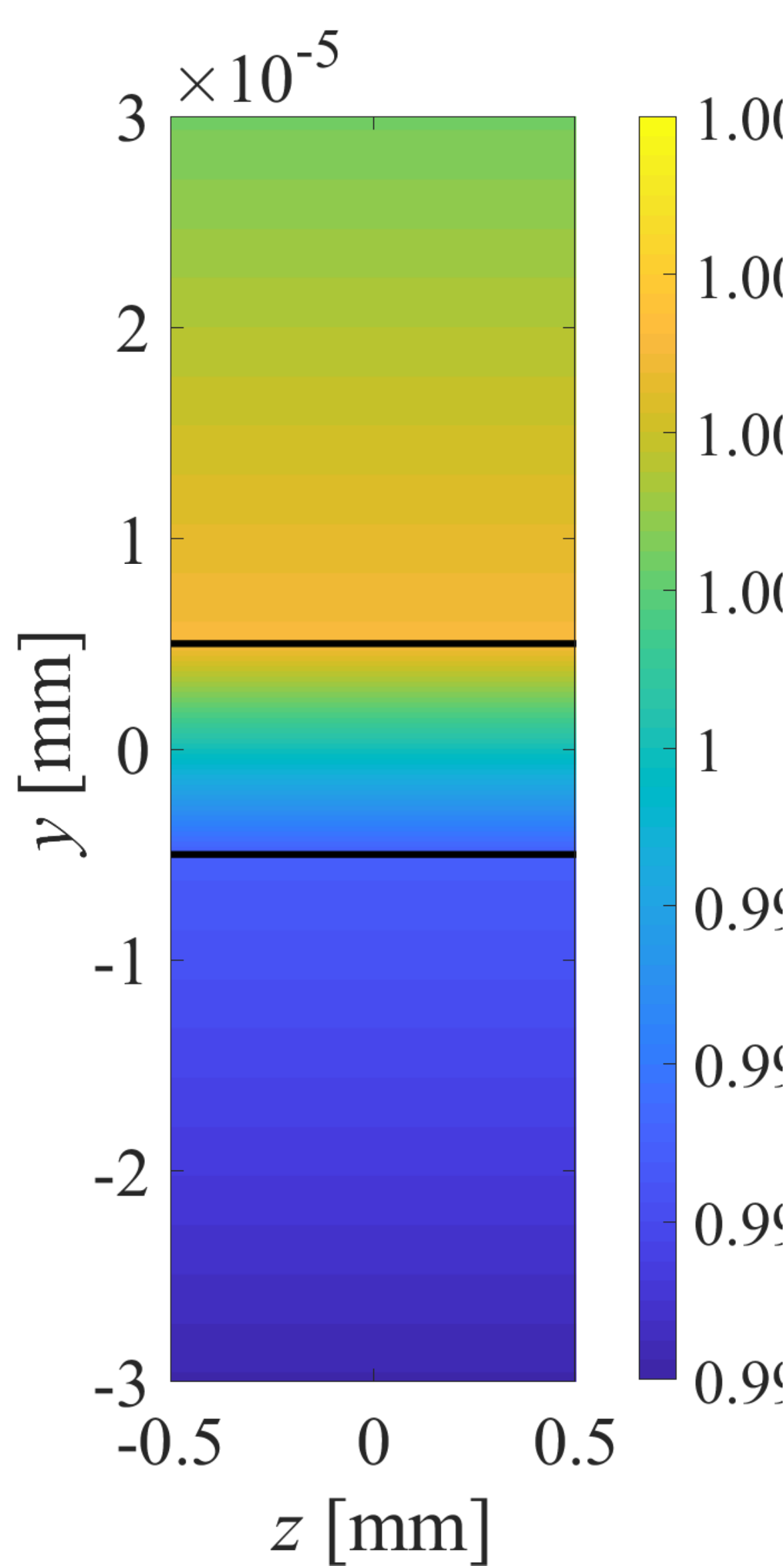}}
	\subfigure[]{\includegraphics[width=0.11\textwidth]{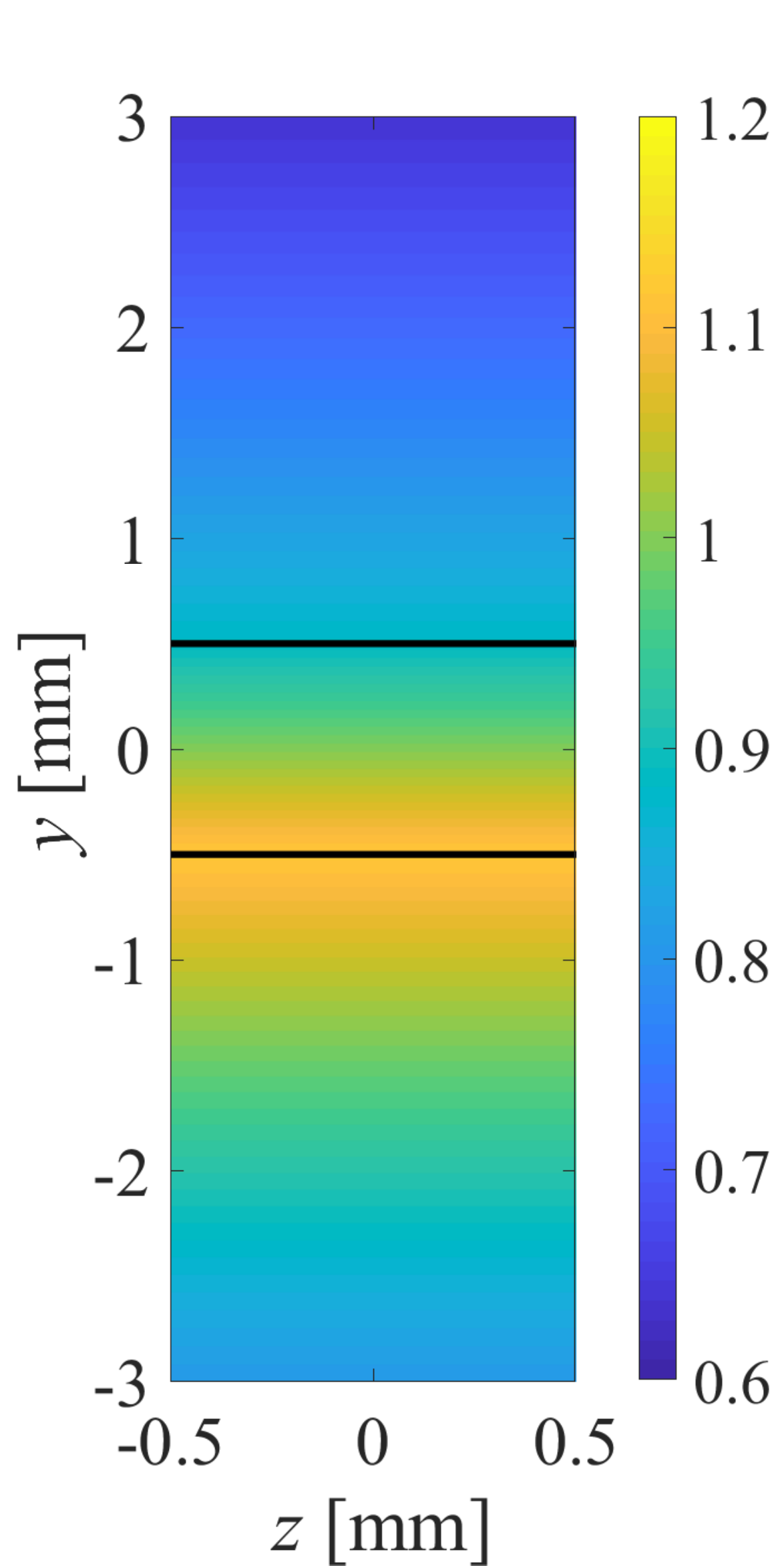}}
	\subfigure[]{\includegraphics[width=0.11\textwidth]{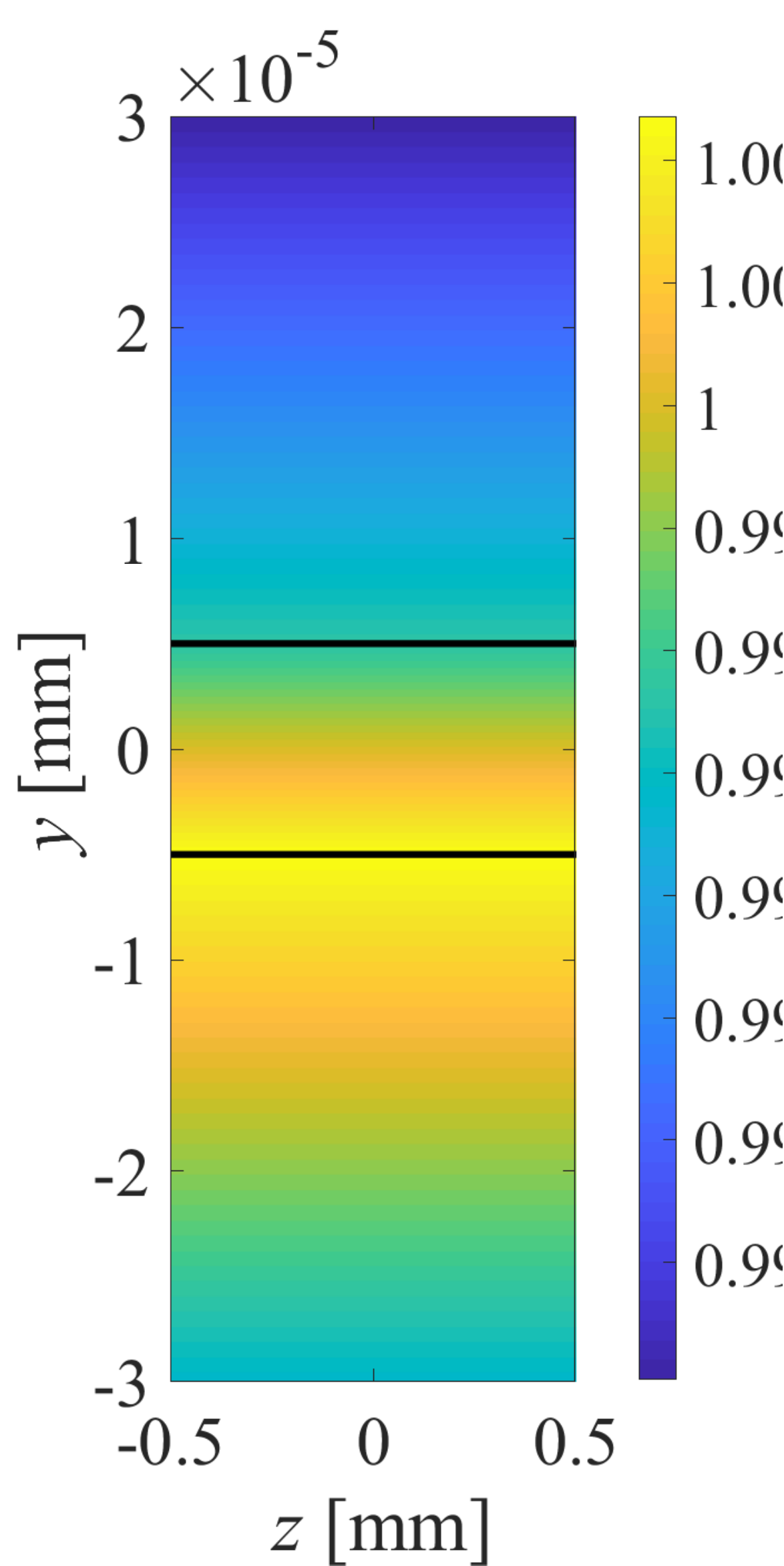}}
	\caption{The distribution of the electric field of two non-resonant sheets. Here, (a) TM polarization, $d=1{~\rm{mm}}$, $\beta=117.4$ 1/m, (b) TM polarization, $d=0.00001{~\rm{mm}}$, $\beta=274$ 1/m, (c) TE polarization, $d=1{~\rm{mm}}$, $\beta=167.7$ 1/m, and (d) TE polarization, $d=0.00001{~\rm{mm}}$, $\beta=153.1$ 1/m. The operational frequency is $f=2.7$ GHz for TM modes and $f=5$ GHz for TE modes. (a) and (b) are the longitudinal component of the electric field. The vertical and horizontal axes are $y$- and $z$-axes, respectively. The metasurface positions are shown by solid black lines.}
	\label{fig:NonResonantfield}
\end{figure} 

\begin{figure*}[ht!]\centering
	\subfigure[]{\includegraphics[width=0.19\textwidth]{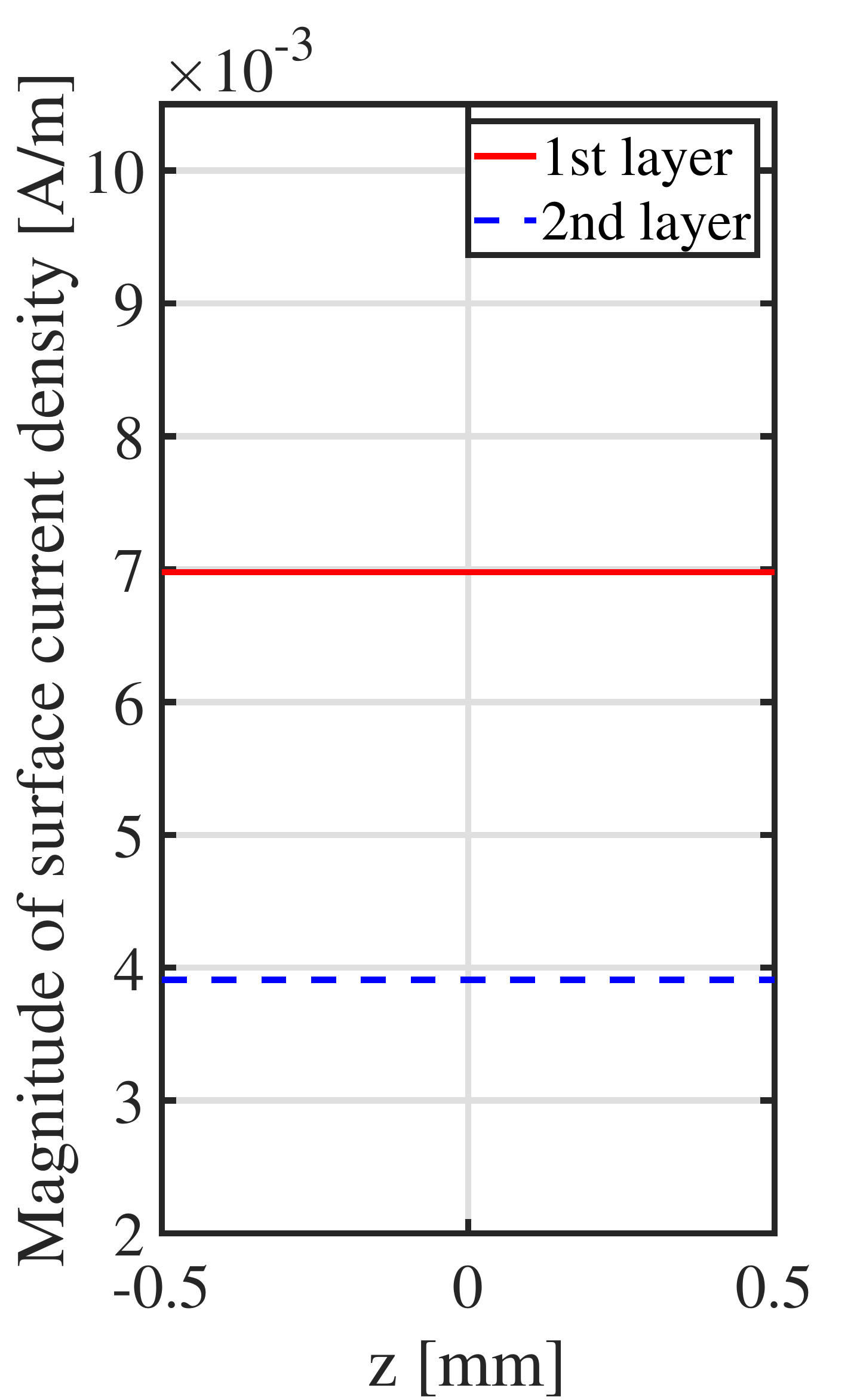}}
	\subfigure[]{\includegraphics[width=0.19\textwidth]{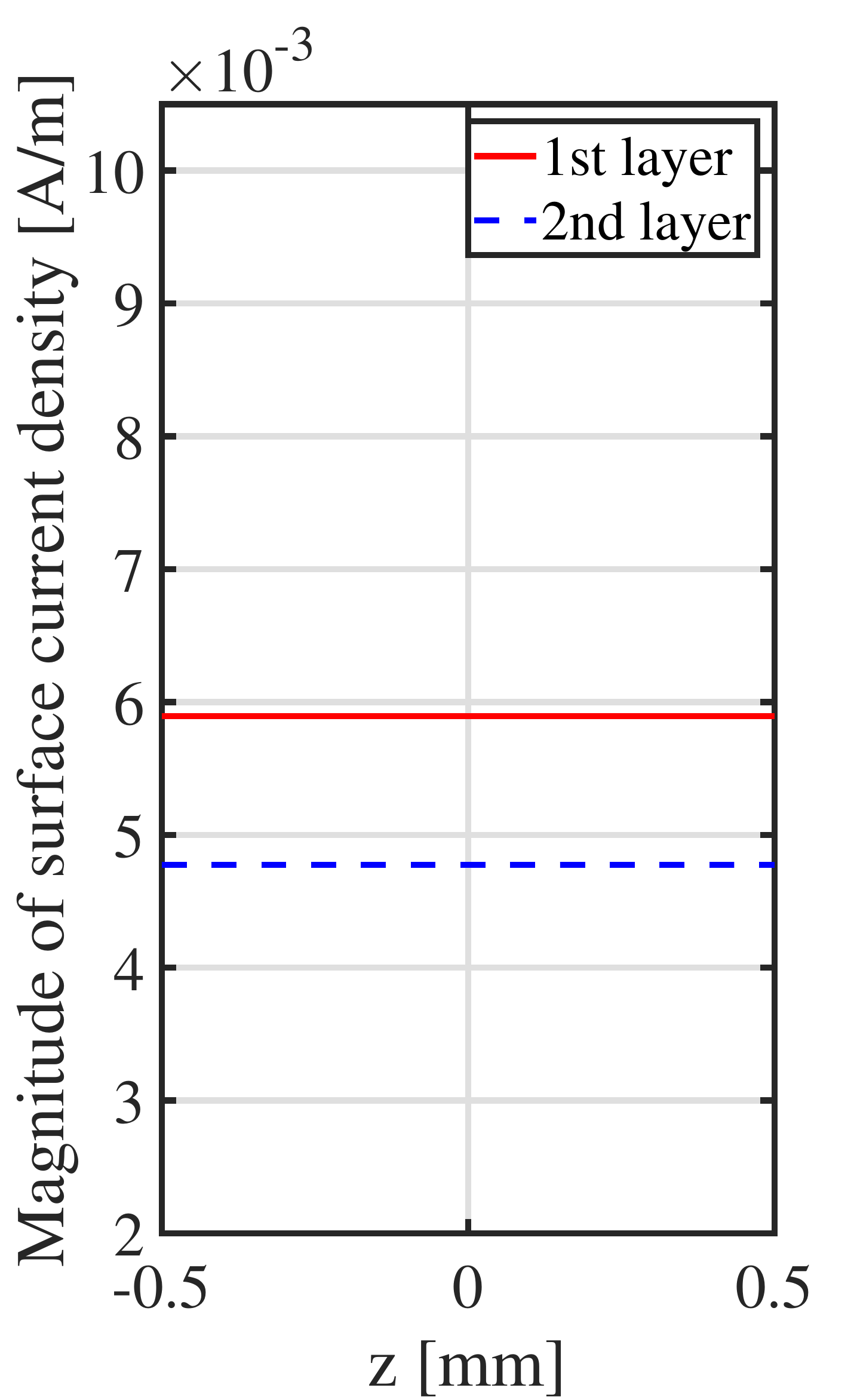}}
	\subfigure[]{\includegraphics[width=0.19\textwidth]{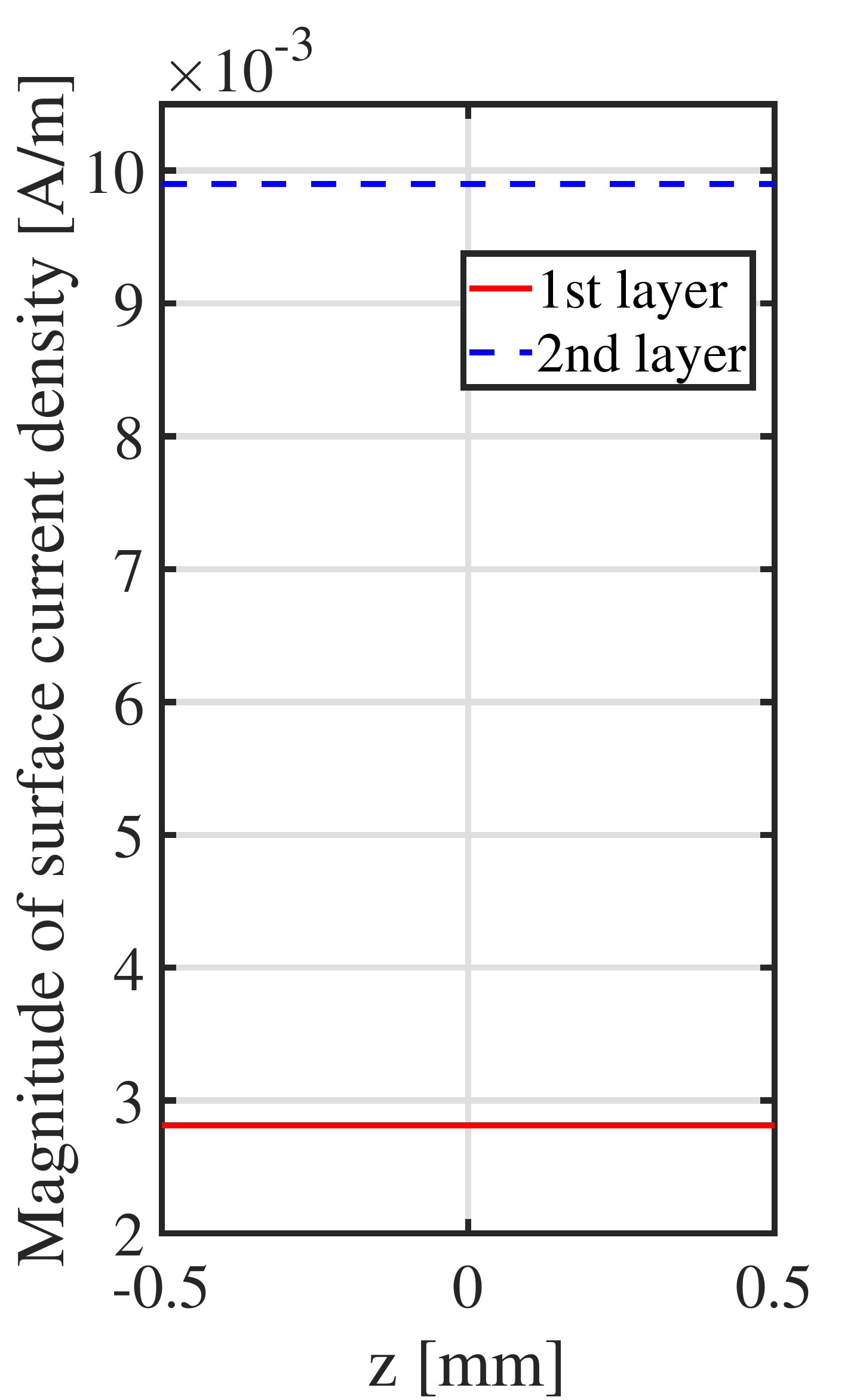}}
	\subfigure[]{\includegraphics[width=0.19\textwidth]{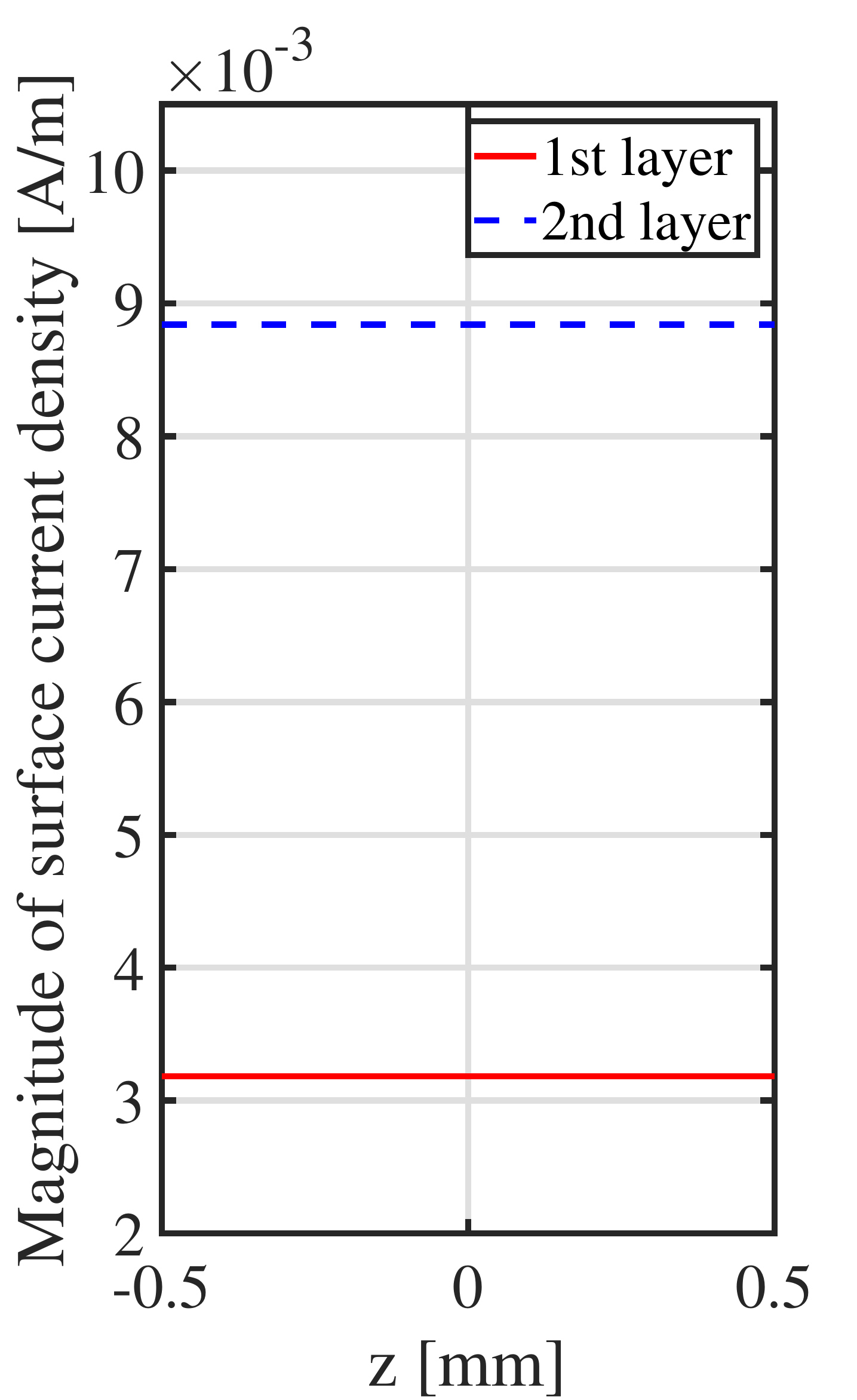}}
	\caption{Magnitude of surface currents for two non-resonant sheets. Here, (a) TM polarization, $d=1{~\rm{mm}}$, $\beta=117.4$ 1/m, (b) TM polarization, $d=0.00001{~\rm{mm}}$, $\beta=274$ 1/m, (c) TE polarization, $d=1{~\rm{mm}}$, $\beta=167.7$ 1/m, and (d) TE polarization, $d=0.00001{~\rm{mm}}$, $\beta=153.1$ 1/m. The operational frequency is $f=2.7$ GHz for TM modes and $f=5$ GHz for TE modes. }
	\label{fig:NonResMagCur}
\end{figure*}

\begin{figure*}[ht!]\centering
	\subfigure[]{\includegraphics[width=0.19\textwidth]{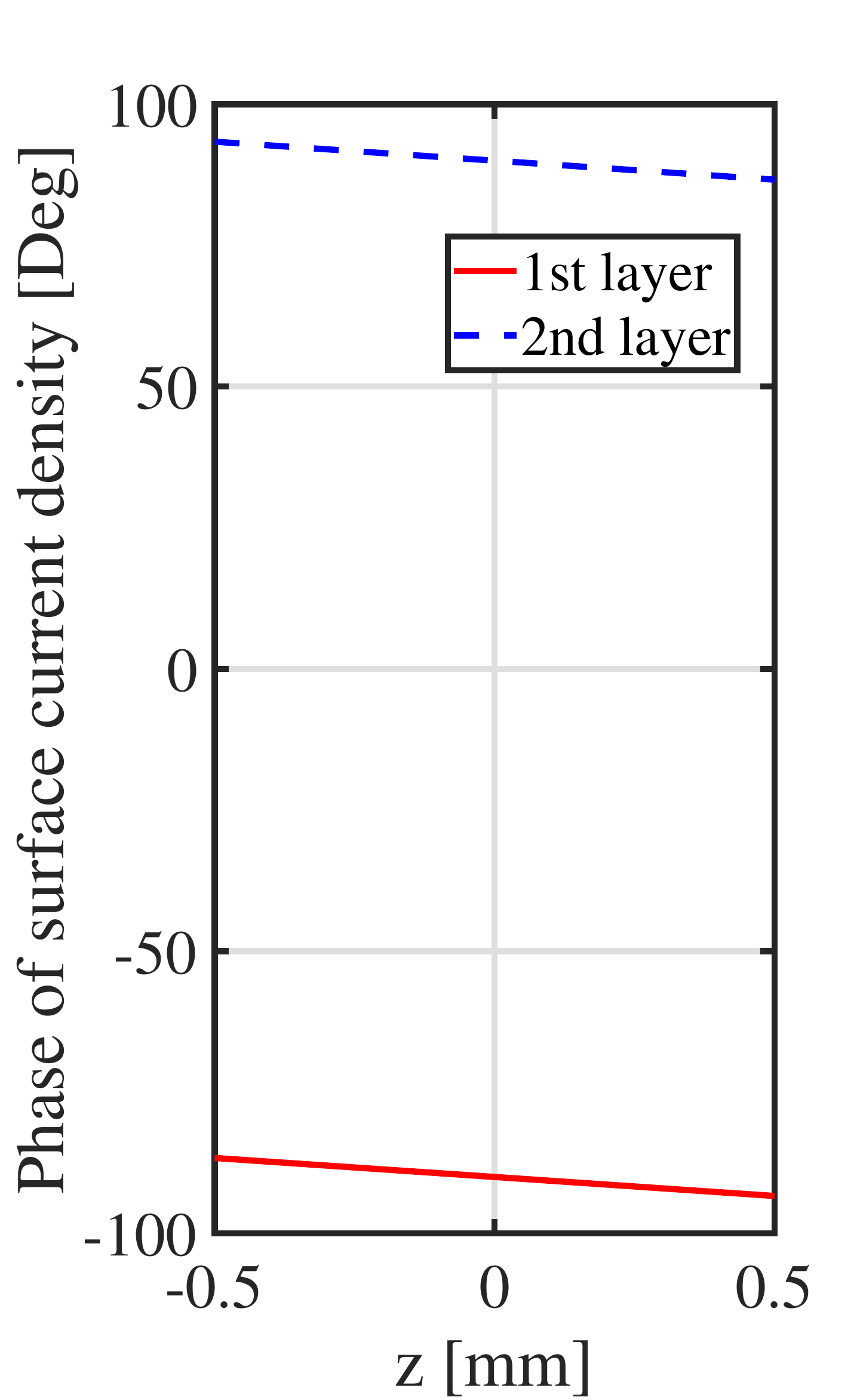}}
	\subfigure[]{\includegraphics[width=0.19\textwidth]{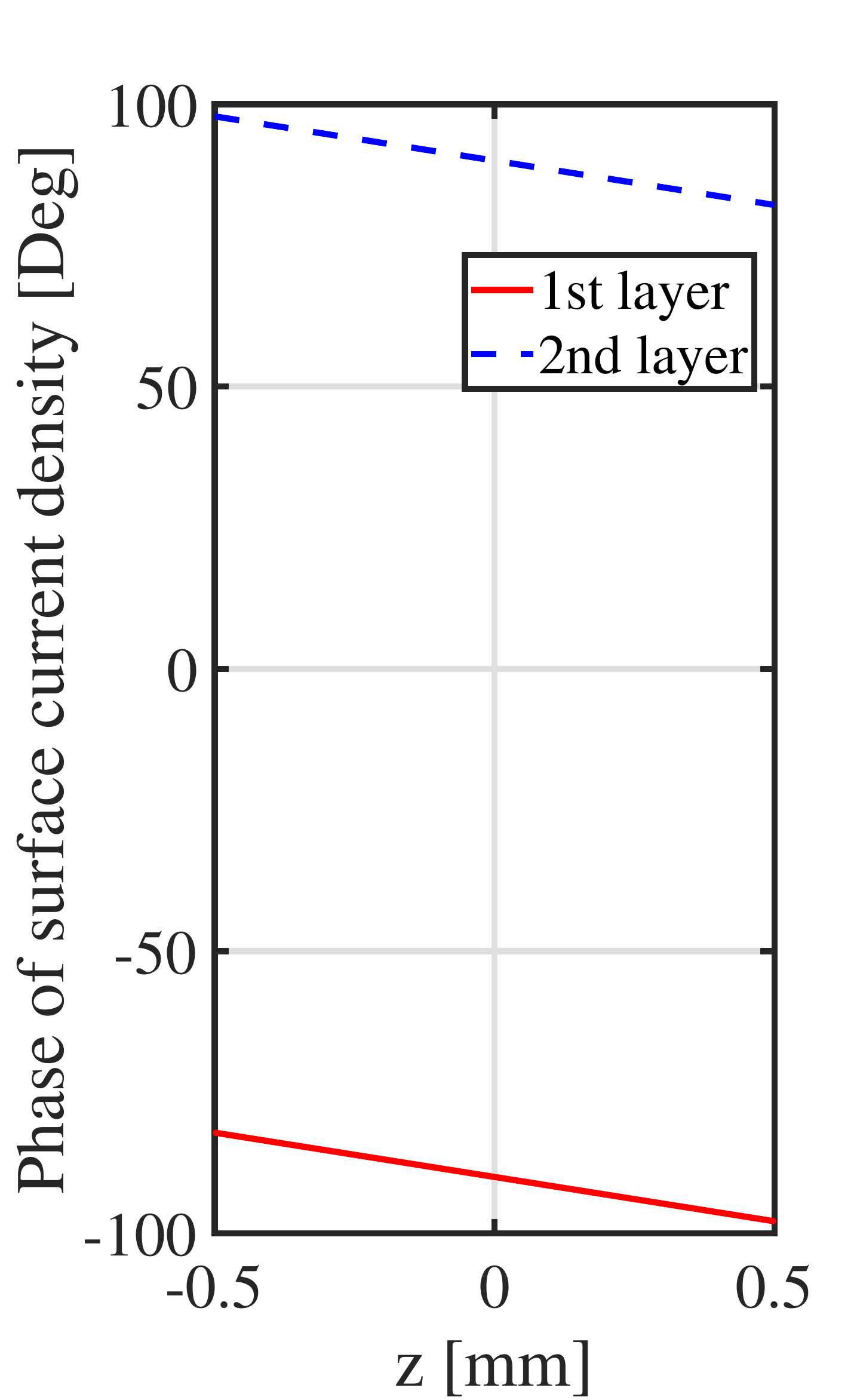}}
	\subfigure[]{\includegraphics[width=0.19\textwidth]{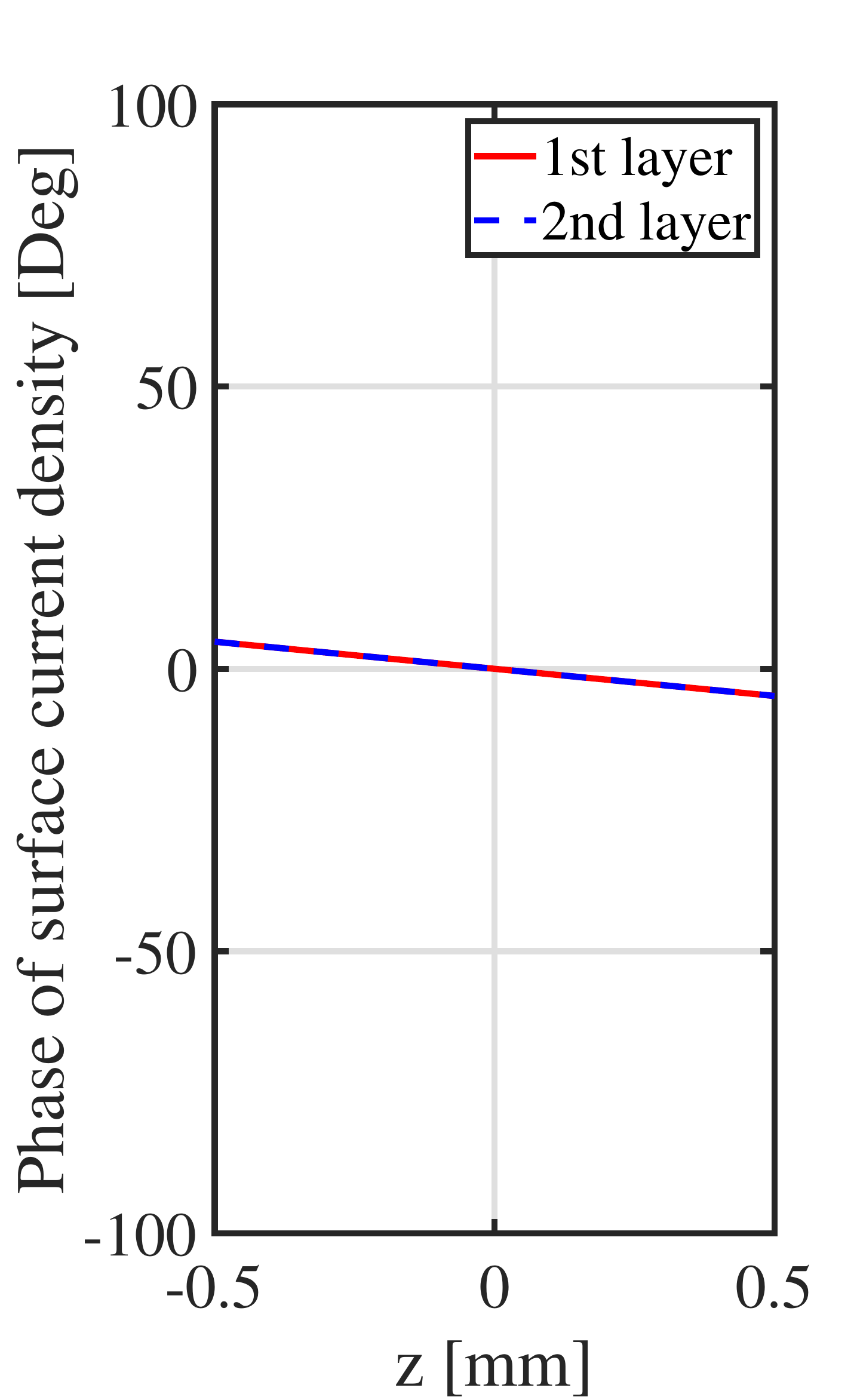}}
	\subfigure[]{\includegraphics[width=0.19\textwidth]{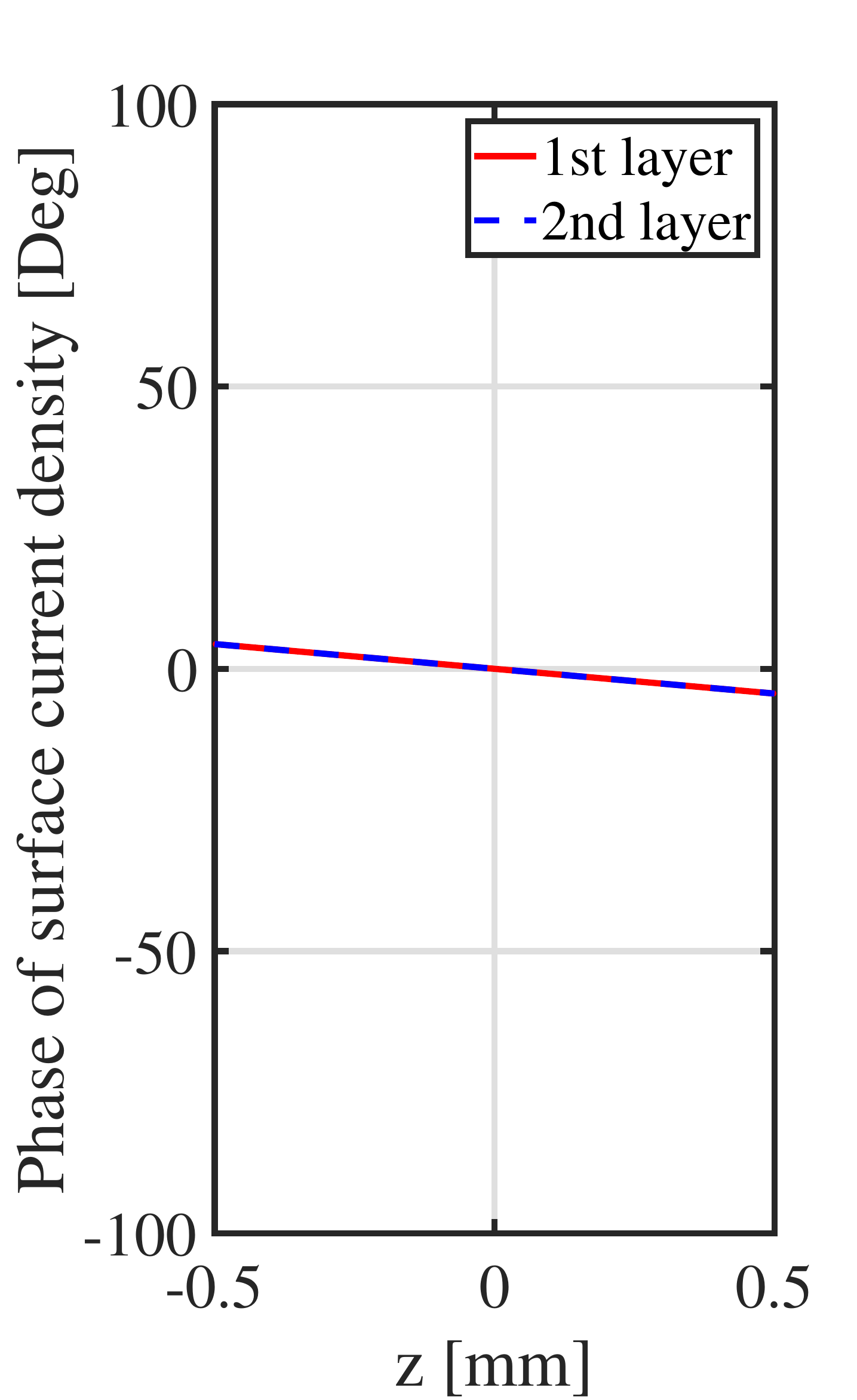}}
	\caption{Phase of surface currents for two non-resonant sheets. Here, (a) TM polarization, $d=1{~\rm{mm}}$, $\beta=117.4$ 1/m, (b) TM polarization, $d=0.00001{~\rm{mm}}$, $\beta=274$ 1/m, (c) TE polarization, $d=1{~\rm{mm}}$, $\beta=167.7$ 1/m, and (d) TE polarization, $d=0.00001{~\rm{mm}}$, $\beta=153.1$ 1/m. The operational frequency is $f=2.7$ GHz for TM modes and $f=5$ GHz for TE modes.}
	\label{fig:NonResPhaCur}
\end{figure*}


\section{Resonant dispersive impedance sheets}
\label{sec:rdis}

In this category, the waveguide under study consists of metasurfaces having resonant properties. In other words, the surface impedance of each metasurface can be expressed as an equivalent impedance of a series or parallel connection of an inductance and capacitance. The unit-cell size of the structure is much smaller than the free-space wavelength (at the resonant frequency),  which allows the structure to be modeled by  homogenized impedances. From the circuit theory point of view, we explicitly know that the equivalent impedance of the series connection of an inductance and capacitance is capacitive and inductive below and above the resonance frequency, respectively. On the contrary, the equivalent impedance of the parallel connection is inductive below the resonance frequency and capacitive above it. Hence, following these considerations, one of the metasurfaces must be realized as a series connection and the other one as a parallel connection of two reactances, as shown in Fig.~\ref{fig:ESresonant}. Here, the surface impedances are characterized by effective inductances and capacitances denoted as $L_1$, $C_1$ and $L_2$, $C_2$ for each metasurface. Let us assume that $L_1$ and $C_1$ are in series connection and $L_2$, $C_2$ are in parallel connection. These values are obtained from the full-wave solution of a plane-wave reflection problem in the quasi-static limit~\cite{self-resonant,self-resonant2} for self-resonant structures. In Fig.~\ref{fig:ESresonant}, we should notice that the corresponding resonance frequencies of two complementary metasurfaces must be the same in order to realize the opposite reactances at all frequencies.

\begin{figure*}[ht!]\centering
	\subfigure[]{\includegraphics[width=0.25\textwidth]{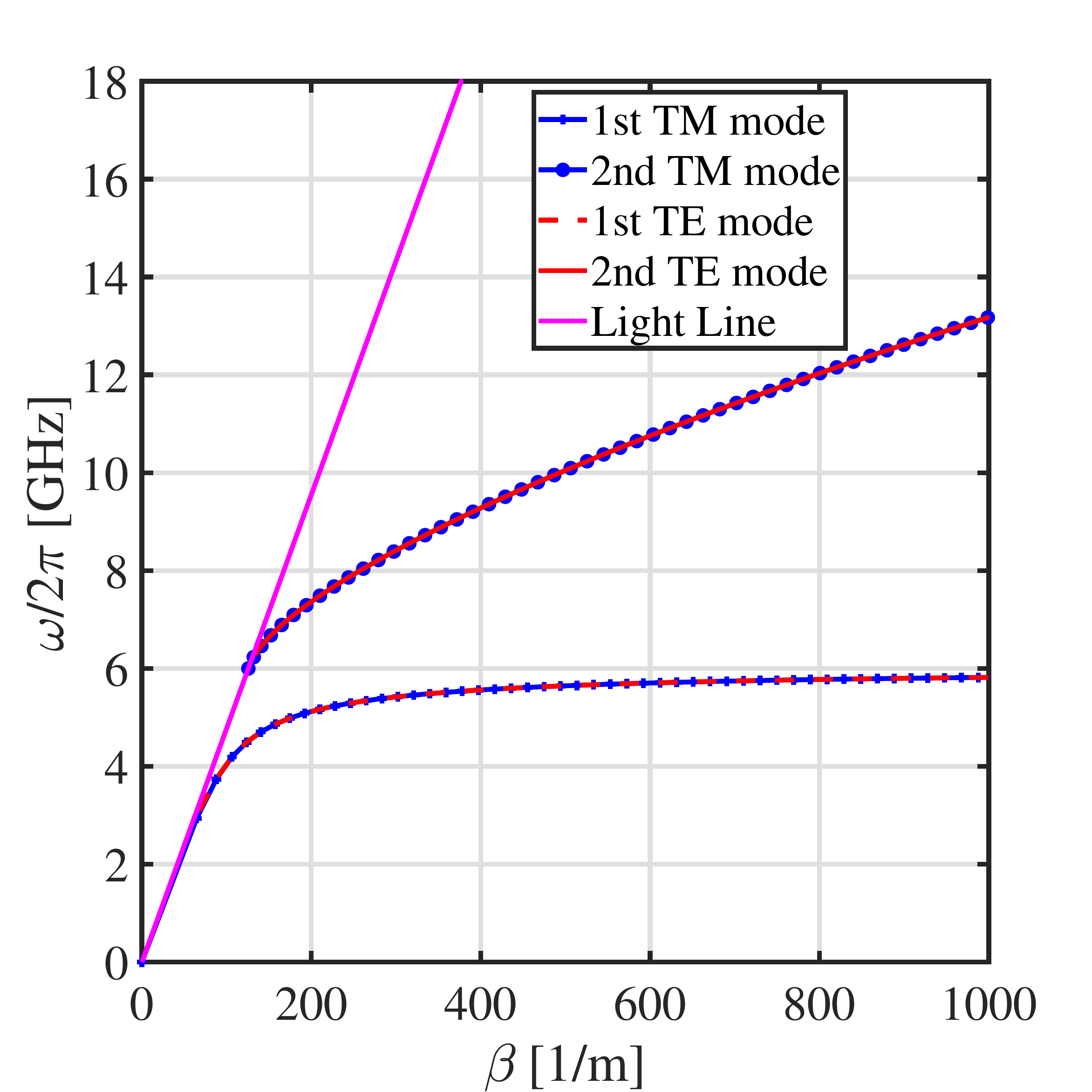}}
	\subfigure[]{\includegraphics[width=0.25\textwidth]{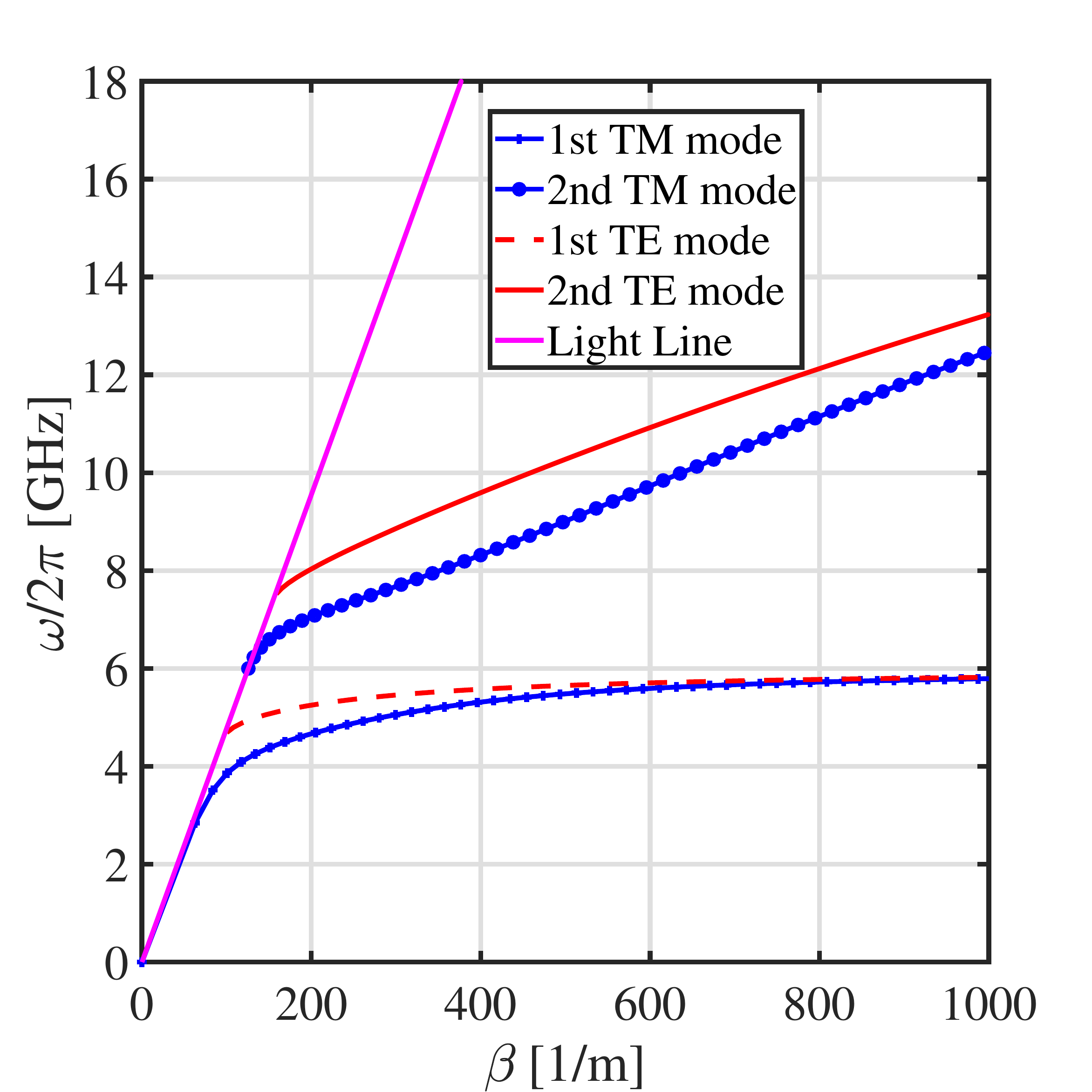}}
	\subfigure[]{\includegraphics[width=0.25\textwidth]{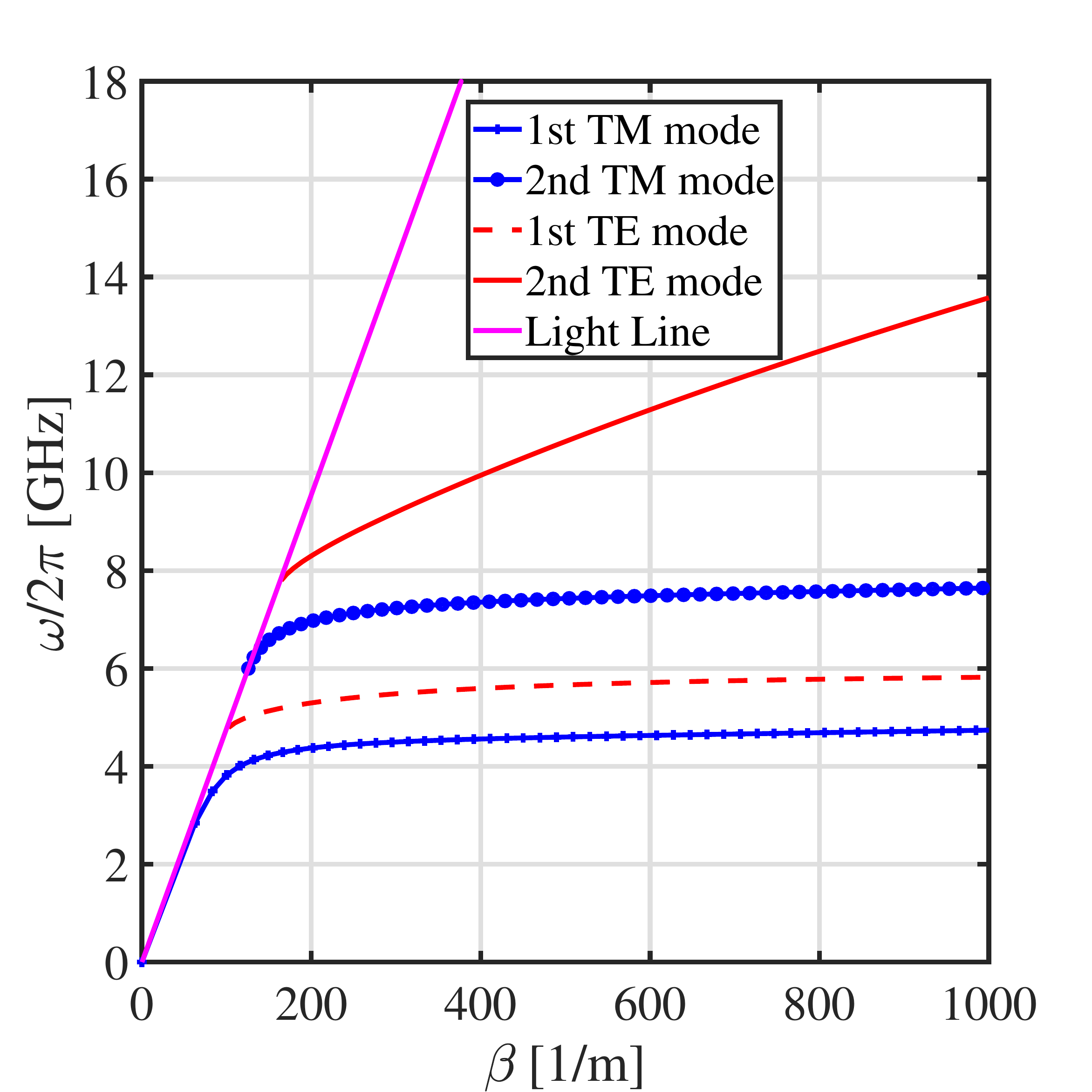}}	
	\caption{Dispersion curves of two resonant sheets, when the distances between the sheets is (a) $d=\lambda_{\rm{6GHz}}$, (b) $d=\lambda_{\rm{6GHz}}/50$, and (c) $d=\lambda_{\rm{6GHz}}/5000$.}
	\label{fig:ResonantDispersion}
\end{figure*}

\begin{figure}[t!]	
\centerline{\includegraphics[width=0.8\columnwidth]{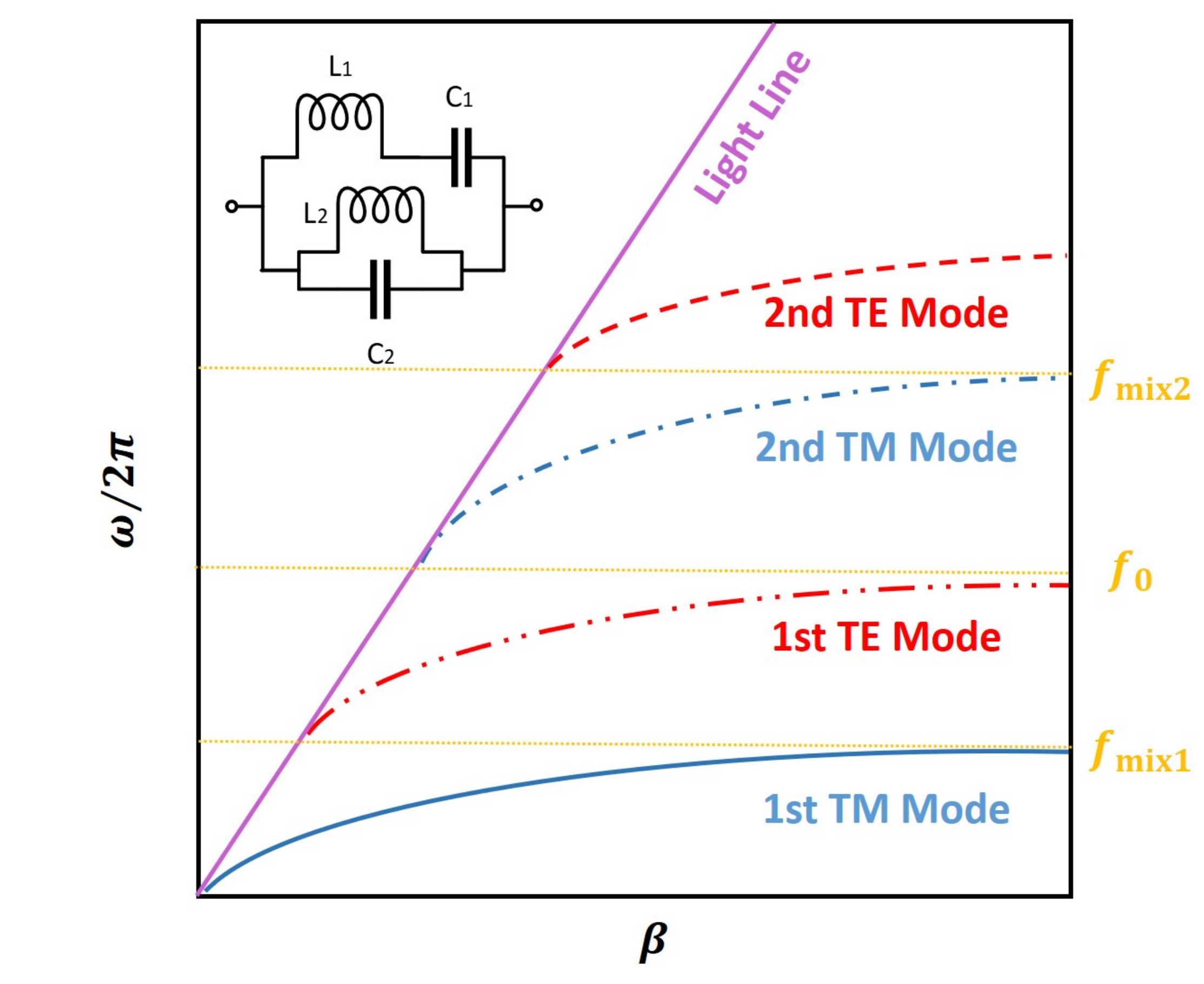}}
\caption{Dispersion curves for two resonant sheets when the distance between  the two sheets tends to zero.}
\label{fig:ResonantExtreme}
\end{figure}

For the two complementary metasurfaces, the impedance of one metasurface is determined by the other metasurface impedance using the Babinet principle (the relations can be found in Eq.~(\ref{eq:LCresonat}) of Appendix). Accordingly, the dispersion relation for two complementary resonant sheets can be written in terms of $L_1$ and $C_1$ as
\begin{equation}
\begin{split}
&2\epsilon_0\alpha+\Big(1-e^{-2\alpha d}\Big)C_1\alpha^2+\Big(2\epsilon_0\alpha+\frac{\epsilon_0\mu_0}{L_1}\Big)\Big(\frac{\omega}{\omega_0}\Big)^4-\cr
&\Big[\frac{\epsilon_0\mu_0}{L_1}+(4+\frac{\eta_0^2C_1}{2L_1})\epsilon_0\alpha+\Big(1-e^{-2\alpha d}\Big)C_1\alpha^2 \Big]\Big(\frac{\omega}{\omega_0}\Big)^2=0,
\end{split}
\end{equation}
for the TM polarization. Analogously, the dispersion relation for the TE polarization reads
\begin{equation}
\begin{split}
&2\mu_0\alpha+\eta_0^2C_1\alpha^2+\Big[2\mu_0\alpha +\frac{\mu_0^2}{L_1}\Big(1-e^{-2\alpha d}\Big)\Big]\Big(\frac{\omega}{\omega_0}\Big)^4-\cr
&\Big[\frac{\mu_0^2}{L_1}\Big(1-e^{-2\alpha d}\Big)+(4+\frac{\eta_0^2C_1}{2L_1})\mu_0\alpha+\eta_0^2C_1\alpha^2)\Big]\Big(\frac{\omega}{\omega_0}\Big)^2=0,\cr
\end{split}
\end{equation}
where $\omega_0$ is the resonant angular frequency of both series and parallel connections. Below/above the resonance frequency, there are always two modes with TE and TM polarizations. Regarding the TM polarization, one of them suffers from a cut-off frequency due to the resonance. For the TE polarization, both of them have cut-off frequencies which can be calculated by (see Eq.~(\ref{eq:fLCresonatcut}) of Appendix)
\begin{equation}
f_{\rm{cut-off}}\approx{f_0\sqrt{\frac{L_1+\frac{\eta_0^2C_1}{8}+\frac{\mu_0d}{2} \pm \sqrt{\bigtriangleup}}{L_1+\mu_0d}}},
\end{equation}
where $\bigtriangleup=(\frac{\eta_0}{2\omega_0})^2+(\frac{\eta_0^2C_1}{8}+\frac{\mu_0d}{2})^2$. As an example, we assume $f_0=6$ GHz as the resonance frequency of resonant metasurfaces, and additionally $L_1=10$~nH and $C_1=0.07$~pF. The values of $L_2$ and $C_2$ are obtained by employing Eq.~(\ref{eq:LCresonat}) of Appendix, which gives $L_2=2.5$~nH and $C_2=0.28$~pF. Figure~\ref{fig:ResonantDispersion} presents the dispersion curves for different distances between the two resonant sheets. As expected, one TE-polarized mode and one TM-polarized mode exist simultaneously below/above the resonance frequency. When $d=\lambda_{\rm{6GHz}}$, the TE and TM modes have approximately the same phase velocity. However, as the distance decreases, the modes are being separated. The limit of $d\rightarrow 0$ brings about two new resonance frequencies $f_{\rm{mix1}}$ and $f_{\rm{mix2}}$ which are given by
\begin{equation}
f_{\rm{mix1}}=\frac{1}{2 \pi}\sqrt{\frac{\eta_0^2C_1+8L_1-\sqrt{\big(\eta_0^2C_1\big)^2+16\eta_0^2C_1L_1}}{8C_1L_1^2}},
\label{eq:fmix1}	
\end{equation}
\begin{equation}
f_{\rm{mix2}}=\frac{1}{2 \pi}\sqrt{\frac{\eta_0^2C_1+8L_1+\sqrt{\big(\eta_0^2C_1\big)^2+16\eta_0^2C_1L_1}}{8C_1L_1^2}}.
\label{eq:fmix2}	
\end{equation}
The dispersion curve of the extreme case is classified by $f_{\rm{mix1}}$, $f_0$ and $f_{\rm{mix2}}$,  which are illustrated in Fig.~\ref{fig:ResonantExtreme}. 

\section{Polarization Insensitivity}
\label{sec:PI}

\begin{figure}[t!]\centering
	\subfigure[]{\includegraphics[width=0.6\columnwidth]{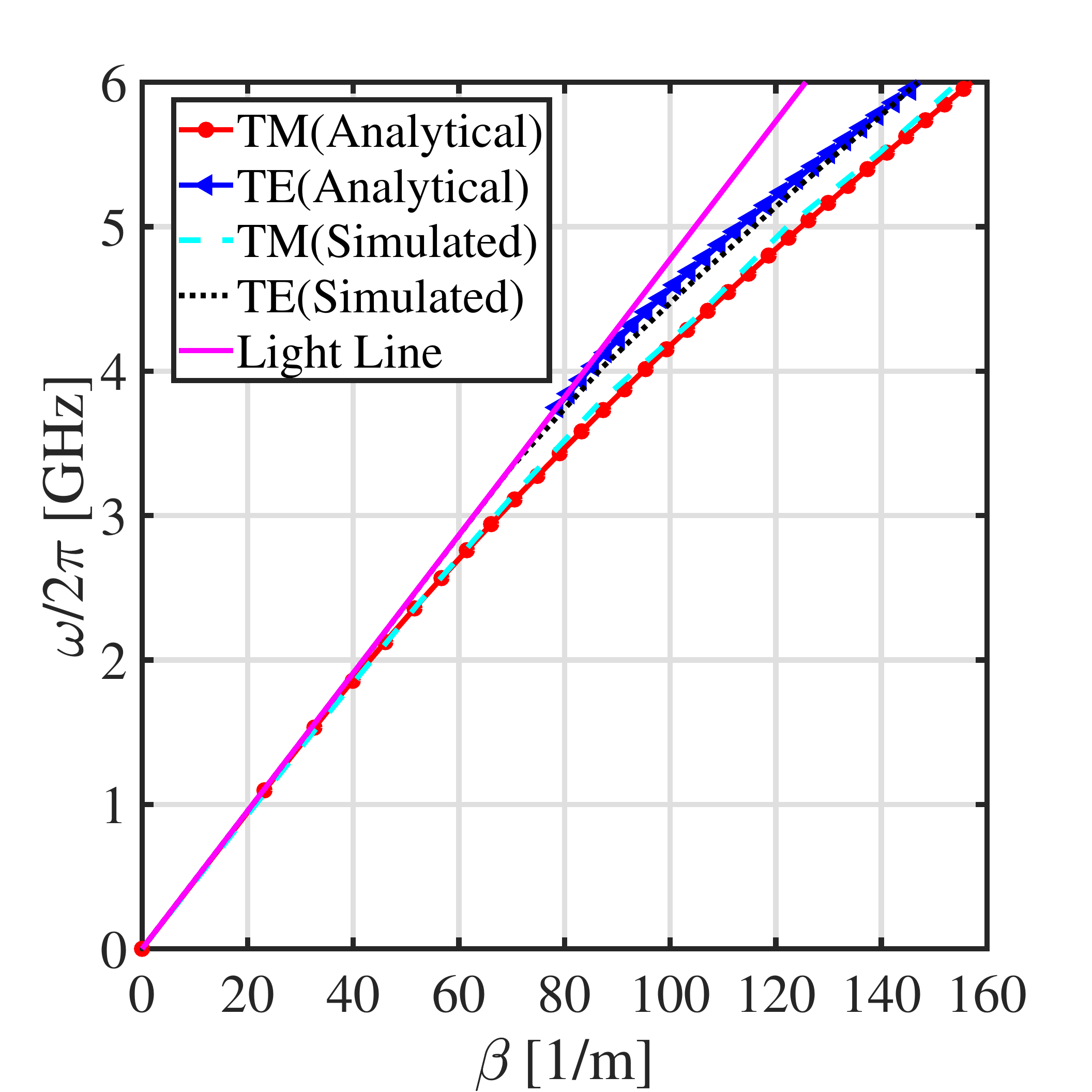}}
	\subfigure[]{\includegraphics[width=0.6\columnwidth]{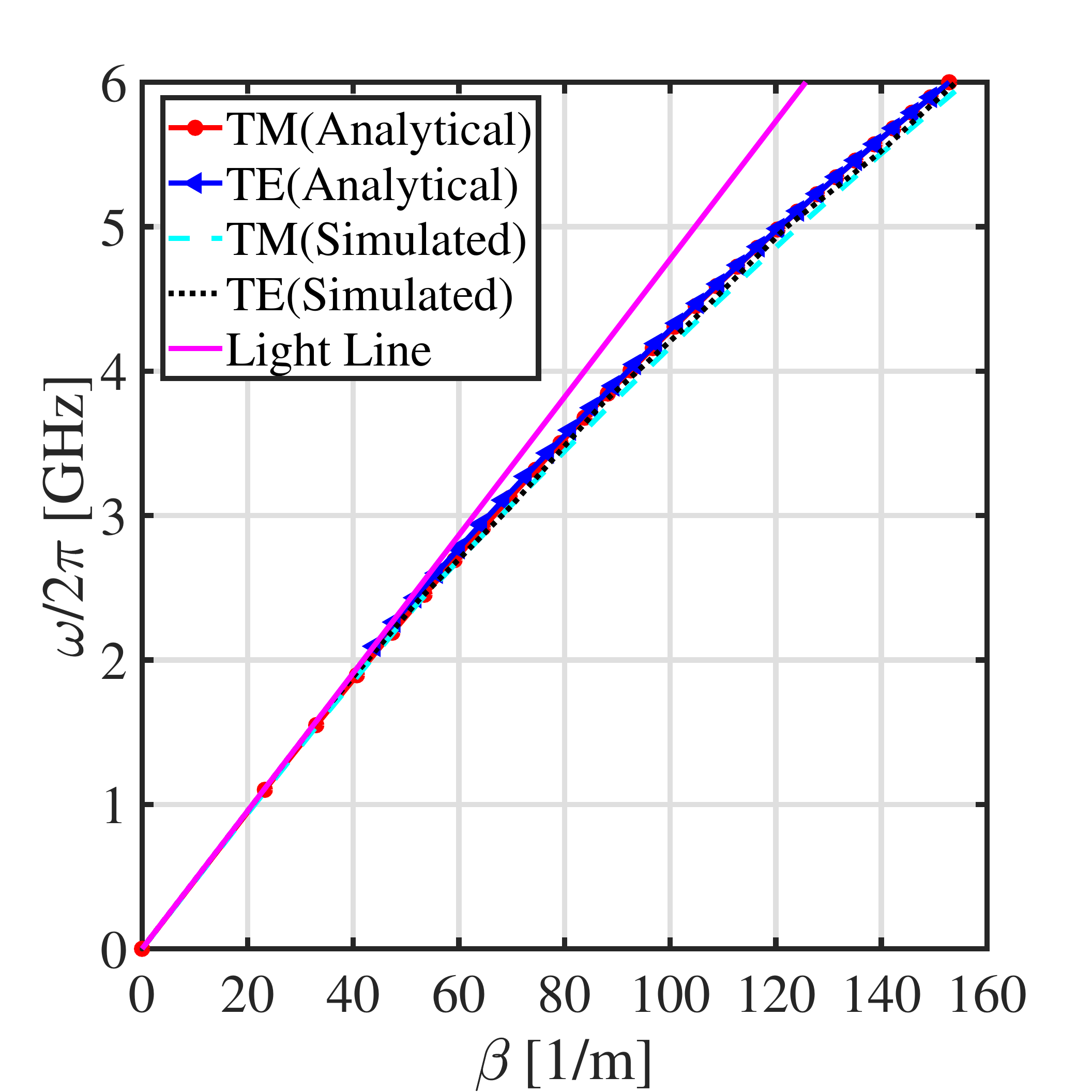}}
	\caption{Dispersion curves of practical realization for the square metal patches and their complementary grid, when the distance between two sheets is: (a) $d=10\, {\rm{mm}}$ and (b) $d=50\, {\rm{mm}}$. Here the length of square patch is 9.8 mm, the width of grid is 0.2 mm and period of unit cell is 10 mm.}
	\label{fig:PracticalRealization10}
\end{figure}
Metasurface-based waveguides have been exploited for the applications of leaky-wave radiation and field focusing~\cite{grbic,holographic}. However, the limitations of applications exist due to polarization sensitivities. Designing a surface-wave waveguide which is insensitive to polarization becomes a significant challenge. Different kinds of waveguide topologies have been explored to propagate waves regardless of polarization at certain frequencies~\cite{Sievenpiper}. Let us define the polarization insensitivity as following: Both TM-polarized and TE-polarized waves can propagate inside the waveguide having the same phase velocity (degeneracy condition). Regarding our proposed waveguide structure, we can find this interesting property based on the dispersion relations written for TE and TM modes. For each value of $\alpha$, the value of $\omega$ is calculated from the general dispersion relations. Subsequently, it is enough to ensure that the solutions are equal to each other (see Appendix~\ref{sec:InseneitiveSupplement}) in order to find the condition which provides us with the property of polarization insensitivity. After some algebraic manipulations, we find the following condition:
\begin{equation}
Z_{\rm{s}1}\cdot Z_{\rm{s}2}=\frac{\eta_0^2}{4}(1-e^{-2\alpha d}).
\label{eq:SamePhaVel}
\end{equation}
However, according to the first equation of the paper, the multiplication of the surface impedances must be equal to $\eta_0^2/4$. Therefore, in the above equation, the expression inside the bracket should be unity meaning that the distance between the two metasurfaces must be considerable compared to the operating wavelength. This was explicitly shown  in Figs.~\ref{fig:NonResonantDispersion} and~\ref{fig:ResonantDispersion}. When the distance between the two metasurfaces is not small, the dispersion curves, corresponding to both modes, match with each other (they have the same phase velocity meaning that the power is transferred regardless of the polarization).

For practical realization,  small square metal patches and their complementary grids are chosen as metasurfaces topology. For different distances, numerical simulations are done using  CST Microwave Studio. The simulation setting is similar to the simulation setting for electromagnetic band gap structures (in which after many numerical simulations and comparing the obtained results with analytical and experimental considerations, a practical rule for choosing background material and boundary setting is found for the correct computation of surface wave dispersion diagram~\cite{Yang}). Regarding the simulations, the key parameter in the computer model is the height of the air space outside the waveguide which emulates  infinite free space outside of the structure (in ``Background Setting'' this space is ``Upper Z-distance'' and ``Lower Z-distance''). Based on numerous computer simulations, an airbox with the height of more than 10 times of the distance between two metasurfaces has to be placed over the unit cell to maintain high accuracy. 

\begin{figure}[t!]\centering
	\subfigure[]{\includegraphics[width=0.6\columnwidth]{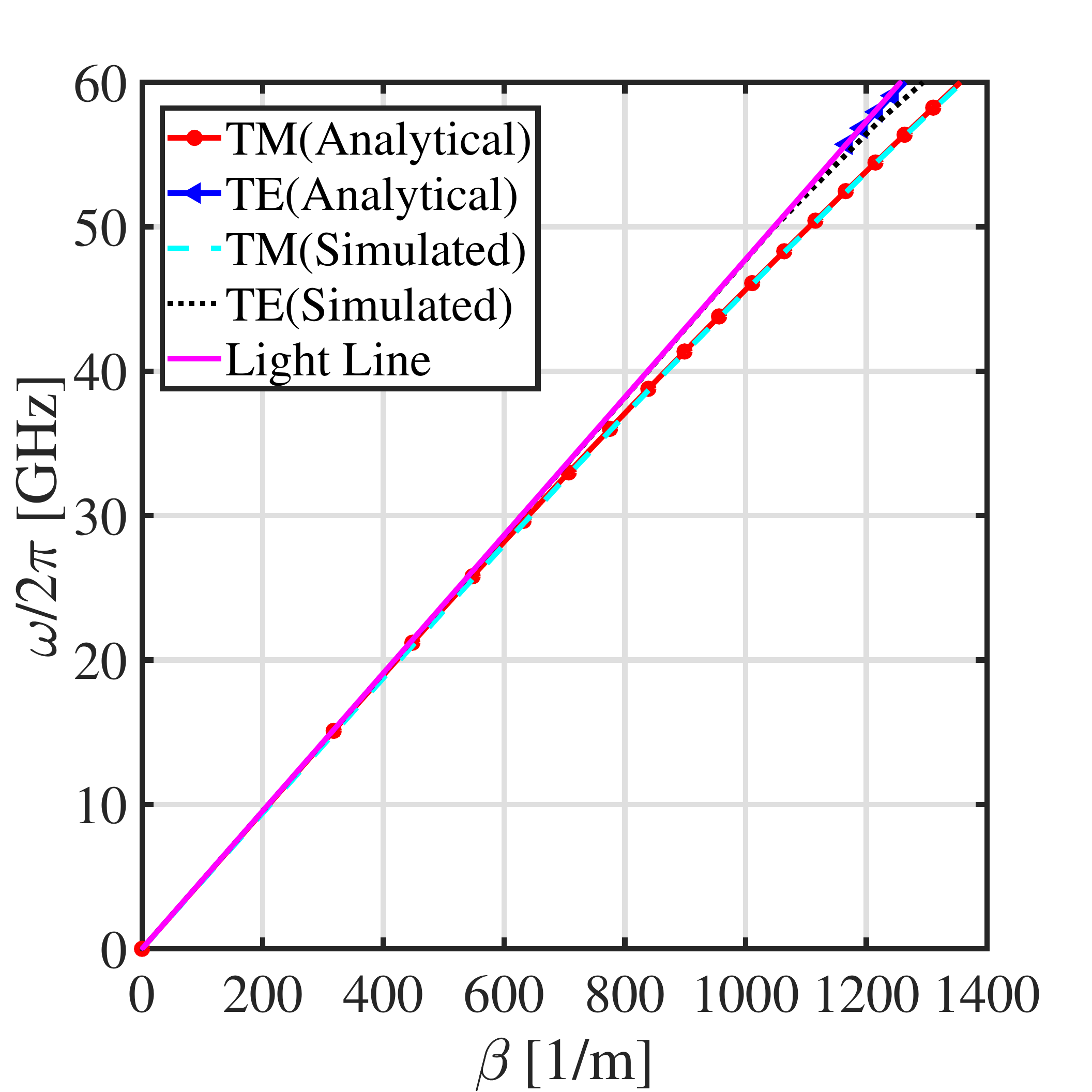}}
	\subfigure[]{\includegraphics[width=0.6\columnwidth]{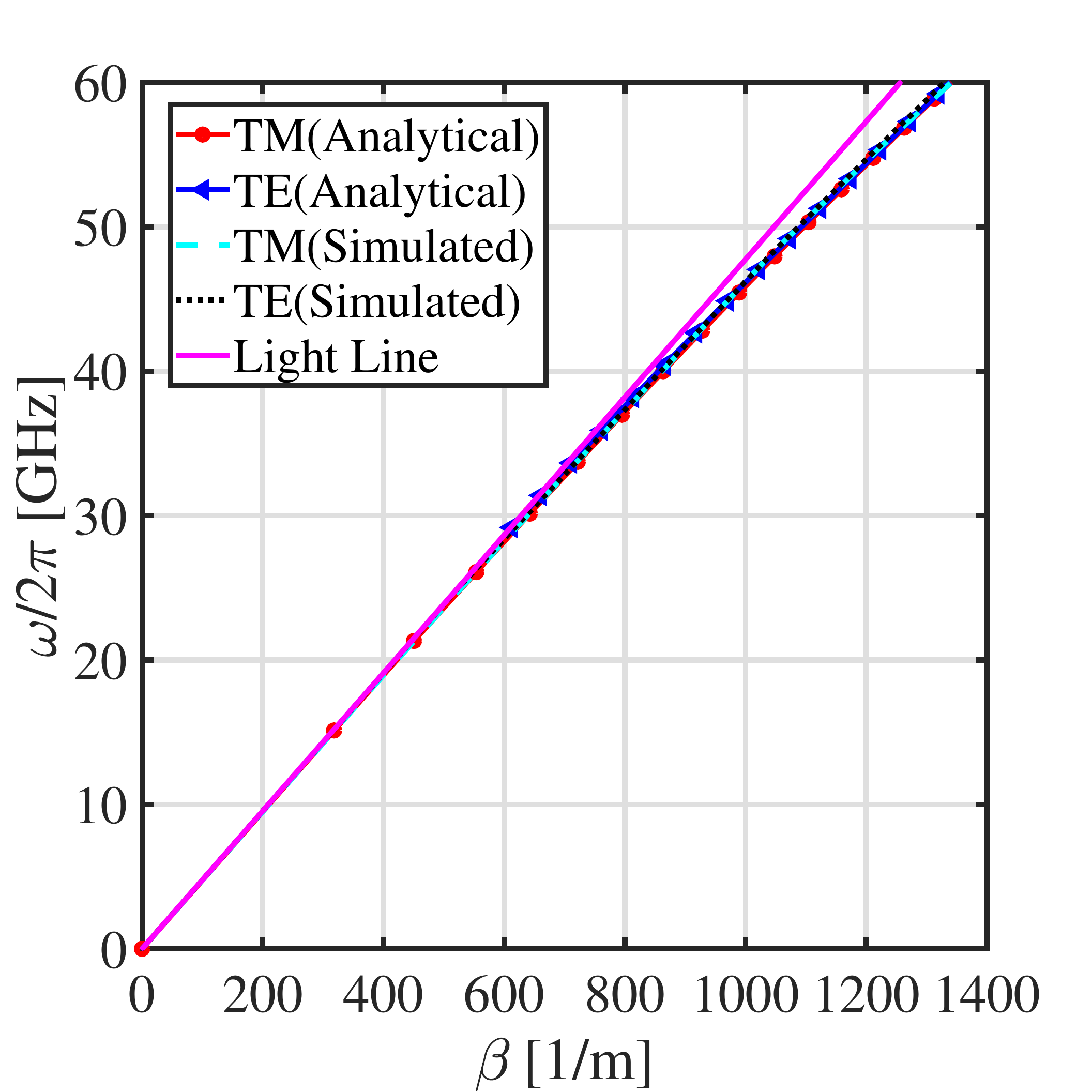}}
	\caption{Dispersion curves of practical realization for the square metal patches and their complementary grid, when the distance between two sheets is: (a) $d=1\, {\rm{mm}}$ and (b) $d=5\, {\rm{mm}}$. Here the length of square patch is 0.9 mm, the width of grid is 0.1 mm and the period is 1 mm.}
	\label{fig:PracticalRealization1}
\end{figure}

Figures~\ref{fig:PracticalRealization10}~and~\ref{fig:PracticalRealization1} show the dispersion curves of complementary metasurfaces with 10 mm and 1 mm unit-cell sizes, respectively. In order to examine the accuracy of the results, in those figures the simulated results are compared with the analytical results. As it is seen, they are in good agreement. One can also see that the dispersion curves of the TM and TE modes overlap in a broad frequency range for larger distances between the two metasurfaces.

\section{Conclusion}
\label{sec:conclusion}

In this paper, we have investigated a waveguiding structure formed by two complementary metasurfaces. We have studied supported guided modes and derived dispersion equations for non-resonant and resonant dispersive impedance sheets. Furthermore, we have investigated the extreme scenarios when the two complementary sheets get close to each other and provided physical insight into the behavior of surface modes. In the limit of zero distance between the two sheets, two new resonance frequencies emerge and four modes are classified by three resonance frequencies. In addition, we have theoretically studied the condition on the sheet impedance to achieve the property of supporting both TM and TE modes with the same phase velocity. Waveguides with such characteristic have great potential to be used in many applications like holographic surfaces or leaky-wave antennas.

\appendix
\section{Capacitance and inductance of non-resonant sheets}
\label{sec:SA}
In our previous work~\cite{xin}, the dispersion relations of a parallel-metasurface waveguide formed by one inductive grid and one capacitive grid are given as
\begin{equation}
\frac{2\epsilon_0\alpha}{C}+\alpha^2(1-e^{-2\alpha d})=\omega^2\Big[\frac{4\epsilon_0^2 L}{C}+2\epsilon_0\alpha L\Big],
\label{eq:TMNonResonant}
\end{equation}
for the TM polarization and
\begin{equation}
\frac{4\alpha^2 L}{C}+\frac{2\mu_0\alpha}{C}=\omega^2\Big[2\mu_0\alpha L+\mu_0^2(1-e^{-2\alpha d})\Big],
\label{eq:TENonResonant}
\end{equation}
for the TE polarization. In this case, the impedances of the two sheets read $Z_{\rm{s}1}= j \omega L$ and $Z_{\rm{s}2}=1/j \omega C$, respectively. Since the two sheets are complementary to each other, we can use the Babinet principle which reads in terms of the sheet impedances ($Z_{\rm{s}1}$ and $Z_{\rm{s}2}$) as~\cite{Tretyakov}
\begin{equation}
Z_{\rm{s}1}\cdot Z_{\rm{s}2}=\frac{\eta_0^2}{4}.
\label{eq:Babinet}
\end{equation}
From the above equation, the relation between $L$ and $C$ for complementary non-resonant sheets is written as
\begin{equation}
C=\frac{4L}{\eta_0^2},
\label{eq:LC}
\end{equation}
Substituting Eq.~(\ref{eq:LC}) into Eqs.~(\ref{eq:TMNonResonant}) and~(\ref{eq:TENonResonant}), we can readily find the dispersion relations for waves along two complementary sheets.
\section{Sheet capacitance and inductance of resonant metasurfaces}
\label{sec:SB}
The dispersion equation of a waveguide formed by one series-connection metasurface and one parallel-connection metasurface can be expressed as~\cite{xin}
\begin{equation}
\begin{split}
&\Big[2\epsilon_0\alpha C_1 C_2 L_1 L_2+4\epsilon_0^2 C_1 L_1 L_2\Big]\omega^4-\cr
&\Big[2\epsilon_0\alpha(C_1 L_1+ C_2 L_2+C_1 L_2)+\alpha^2\Big(1-e^{-2\alpha d}\Big)C_1C_2 L_2   +\cr
&4\epsilon_0^2L_2\Big]\omega^2+2\epsilon_0\alpha+\alpha^2\Big(1-e^{-2\alpha d}\Big)C_1=0,
\label{eq:TMResonant}
\end{split}
\end{equation}
for TM-mode waves and
\begin{equation}
\begin{split}
&\Big[2\mu_0\alpha C_1C_2L_1L_2+\mu_0^2\Big(1-e^{-2\alpha d}\Big)L_2C_1C_2\Big]\omega^4-\cr
&\Big[4\alpha^2C_1L_1L_2+2\mu_0\alpha ( C_1L_1+C_2L_2+C_1L_2)+\cr
&\mu_0^2\Big(1-e^{-2\alpha d}\Big)C_1\Big]\omega^2+4\alpha^2 L_2+2\mu_0\alpha=0,
\label{eq:TEResonant}
\end{split}
\end{equation}
 for TE-mode waves. The series-connection impedance can be represented as $Z_{\rm{s}1}=j\omega L_1+1/j \omega C_1$ and parallel-connection impedance is written as $Z_{\rm{s}2}=j\omega L_2/(1-\omega^2 L_2C_2)$. Two complementary resonant sheets must have the same resonant angular frequency denoted as
\begin{equation}
\omega_0=\frac{1}{\sqrt{L_1C_1}}=\frac{1}{\sqrt{L_2C_2}}.
\end{equation}
According to the Babinet principle (Eq.~(\ref{eq:Babinet})), the effective inductance and capacitance of the parallel-connection impedance are related to the inductance and capacitance of series-connection impedance as
\begin{equation}
L_2=\frac{\eta_0^2}{4}C_1,  C_2=\frac{4L_1}{\eta_0^2}
\label{eq:LCresonat}
\end{equation}
Substituting Eq.~(\ref{eq:LCresonat}) into Eqs.~(\ref{eq:TMResonant}) and~(\ref{eq:TEResonant}), the dispersion relations for complementary sheets can be obtained. 

Interestingly, both  TE-polarized modes suffer from cut-off frequencies when the attenuation constant $\alpha$ approaches zero, which can be approximately evaluated as
\begin{equation}
f_{\rm{cut-off}}\approx{f_0\sqrt{\frac{1}{4(L_1+\mu_0d)}\Big[4L_1+\frac{\eta_0^2C_1}{2}+2\mu_0d \pm \sqrt{\bigtriangleup} \Big]}}
\label{eq:fLCresonatcut}
\end{equation}
where $\bigtriangleup=(4L_1+\frac{\eta_0^2C_1}{2}+2\mu_0d)^2-16L_1^2-16\mu_0dL_1$, which can be simplified as $\bigtriangleup=4\eta_0^2L_1C_1+(\frac{\eta_0^2C_1}{2}+2\mu_0d)^2$. Obviously, the expression of Eq.~(\ref{eq:fLCresonatcut}) under the two square roots on the right-hand side are always non-negative, which results in two real solutions.
\section{Condition for achieving the same phase velocity of both modes}
\label{sec:InseneitiveSupplement}
Based on the definition of phase velocity,  the ratio of the angular frequency and the phase constant should be the same at each value of $\alpha$ for TE- and TM-polarized waves when the phase velocities of TE- and TM-polarized waves are same. We denote this ratio by 
\begin{equation}
\frac{\omega_{\rm{TE}}}{\beta_{\rm{TE}}}=\frac{\omega_{\rm{TM}}}{\beta_{\rm{TM}}}=A(\alpha).
\end{equation}
Here, $\omega_{\rm{TE(TM)}}$ and $\beta_{\rm{TE(TM)}}$ are the angular frequencies and the phase constants for the TE and TM polarizations, respectively. Based on the relation between $\omega$, $\alpha$, and $\beta$ for TE and TM polarizations,  we can find that 
\begin{equation}
\alpha^2=\omega_{\rm{TE}}^2\big[\frac{1}{A(\alpha)^2}-\frac{1}{c^2}\big],
\label{eq:AlfaOmegaTE}
\end{equation}
\begin{equation}
\alpha^2=\omega_{\rm{TM}}^2\big[\frac{1}{A(\alpha)^2}-\frac{1}{c^2}\big],
\label{eq:AlfaOmegaTM}
\end{equation}
where $c$ is the speed of light.
Comparing the above two equations, the expressions in square brackets on the right-hand side are the same for each value of $\alpha$. Therefore, it can be concluded that the angular frequencies of waves of TE and TM polarizations at each value of $\alpha$ should be the same.

In~\cite{xin}, the dispersion relations for the waveguides which have arbitrary sheet impedance have been derived: 
\begin{equation}
\alpha^2\Big(e^{-2\alpha d}-1\Big)=j2\omega\epsilon_0\Big(Z_{\rm{s}1}+Z_{\rm{s}2}\Big)\alpha-4\omega^2\epsilon_0^2Z_{\rm{s}1}Z_{\rm{s}2},
\label{eq:gedisrelz3z2} 
\end{equation}
for TM polarization and 
\begin{equation}
4Z_{\rm{s}1}Z_{\rm{s}2}\alpha^2+j2\omega\mu_0(Z_{\rm{s}1}+Z_{\rm{s}2})\alpha+\omega^2\mu_0^2 (e^{-2\alpha d}-1)=0,
\label{eq:DisR}
\end{equation}
for TE polarization. At each value of $\alpha$, the value of $\omega$ calculated from Eqs.~(\ref{eq:gedisrelz3z2}) and~(\ref{eq:DisR}) should be the same. Using this  condition, after some algebraic manipulations, we come to  Eq.~(\ref{eq:SamePhaVel}). 

%

\end{document}